\documentclass{llncs}
\usepackage{makeidx}  
\usepackage{url}
\usepackage{afterpage}
\usepackage{amssymb,amsfonts}
\usepackage{yfonts} 
\usepackage{caption}

\newcommand{\mc}{\textnormal{MC}}
\newcommand{\dtmc}{\textnormal{MC}}
\newcommand{\imc}{\textnormal{IMC}}
\newcommand{\pimc}{\textnormal{pIMC}}
\newcommand{\pmc}{\textnormal{pMC}}

\newcommand{\csp}{\textnormal{CSP}}

\newcommand{\Dist}{\ensuremath{\mathsf{Dist}}}

\newcommand{\mcSet} {\ensuremath{\tt{MC}}}
\newcommand{\pmcSet}{\ensuremath{\tt{pMC}}}
\newcommand{\imcSet}{\ensuremath{\tt{IMC}}}
\newcommand{\pimcSet}{\ensuremath{\tt{pIMC}}}
\newcommand{\aminstance}{\ensuremath{{\tt L}}}

\newcommand{\satisfaction}{\ensuremath{\models}}
\newcommand{\satisfactionImcOnce}{\ensuremath{\models^{o}_{\tt{I}}}}
\newcommand{\satisfactionImc}{\ensuremath{\models^{a}_{\tt{I}}}}
\newcommand{\satisfactionPmc}{\ensuremath{\models_{\tt{p}}}}
\newcommand{\satisfactionPimc}{\ensuremath{\models^a_{\tt{pI}}}}
\newcommand{\satisfactionPimcOnce}{\ensuremath{\models^o_{\tt{pI}}}}

\newcommand{\entailed}{\ensuremath{\sqsubseteq}}

\newcommand{\Mec} {\ensuremath{{\bf C_{{\exists}c}}}}
\newcommand{\Mer}{\ensuremath{{\bf C_{{\exists}r}}}}

\newcommand{\MerPrime}{\ensuremath{{\bf C^{\prime}_{{\exists}r}}}}
\newcommand{\MerExt}{\ensuremath{{\bf C_{{\exists}\bar{r}}}}}

\newcommand{\ie} {{\em i.e.},\ }
\newcommand{\eg} {{\em e.g.},\ }
\newcommand{\cf} {cf.\ }

\newcommand{\ttransition}[1]{\ensuremath{\theta_{#1}}}

\newcommand{\transition}[2]{\ensuremath{\ttransition{#1}^{#2}}}

\newcommand{\transitionSet}{\ensuremath{\Theta}}

\newcommand{\Succ}{\ensuremath{{\tt{Succ}}}}
\newcommand{\Pred}{\ensuremath{{\tt{Pred}}}}

\newcommand{\Inter}  {\ensuremath{{\mathbb{I}}}}

\newcommand{\bench}[1]    {\textnormal{\textsc{#1}}}
\newcommand{\nsatisfaction}[3] {\ensuremath{#2 \oset[-0.5ex]{\text{\hspace*{0.02cm}\small{#1}}}{\satisfactionImc} #3}}

\newcommand{\pctlStar} {PCTL$^*$}

\newcommand{\Nset}             {\ensuremath{\mathbb{N}}}
\newcommand{\Rset}             {\ensuremath{\mathbb{R}}}
\newcommand{\Qset}             {\ensuremath{\mathbb{Q}}}
\newcommand{\Proba}            {\ensuremath{\mathbb{P}}}

\newcommand{\true}             {\ensuremath{\mathsf{true}}}
\newcommand{\false}            {\ensuremath{\mathsf{false}}}

\newcommand{\ssum}[2]{\sum\limits_{\mathclap{#1}^{#2}}}

\newcommand{\ltlUntil}{\textnormal{ U }}

\newcommand{\ltlExists}{\ensuremath{\Diamond}}

\newcommand{\st}{\ensuremath{\ | \ }}

\usepackage[dvipsnames]{xcolor}
\usepackage{varwidth}
\definecolor{light-gray}{gray}{0.85}

\newcommand{\custompar}[1]{\smallskip \noindent {\bf #1}}

\newcommand{\enumerateConstraints}{
	\renewcommand{\labelenumi}{\theenumi}
	\renewcommand{\theenumi}{(\textbf{\arabic{enumi})}}}

\usepackage{etoolbox}
\makeatletter
\patchcmd{\@addmarginpar}{\ifodd\c@page}{\ifodd\c@page\@tempcnta\m@ne}{}{}
\makeatother
\reversemarginpar
\usepackage{mathtools}

\usepackage[normalem]{ulem} 
\usepackage[inline]{enumitem} 

\makeatletter
\newcommand{\oset}[3][0ex]{%
  \mathrel{\mathop{#3}\limits^{
    \vbox to#1{\kern-2\ex@
    \hbox{$\scriptstyle#2$}\vss}}}}
\makeatother

\newlist{customenum}{enumerate}{1}
\setlist[customenum,1]{%
  label=(\arabic*),
}
\usepackage{tikz}
\usetikzlibrary{arrows}
\tikzset{
	vertex/.style   = {shape=circle,draw,minimum size=1em, font=\tiny},
	selected/.style = {thick},
	covered/.style  = {draw=blue,blue},
	edge/.style = {->,> = latex', font=\tiny}
}

\begin{document}

\title{Reachability in Parametric Interval Markov Chains using Constraints}
%
%

\author{Anicet Bart\inst{1} \and Beno\^it Delahaye\inst{2} \and
Didier Lime\inst{3} \and \\
\'Eric Monfroy\inst{2} \and Charlotte Truchet\inst{2}}
\authorrunning{Anicet Bart et al.} 
%
\tocauthor{Anicet Bart, Beno\^it Delahaye, Didier Lime, 
\'Eric Monfroy and Charlotte Truchet}

\institute{
Institut Mines-T\'el\'ecom Atlantique - LS2N, UMR 6004 - Nantes, France
\and
Universit\'e de Nantes - LS2N, UMR 6004 - Nantes, France
\and
\'{E}cole Centrale de Nantes - LS2N, UMR 6004 - Nantes, France
\email{<firstname>.<lastname>@ls2n.fr}
}

  \maketitle              

\begin{abstract}

Parametric Interval Markov
  Chains ({\pimc}s) are a specification formalism that extend Markov
  Chains ({\mc}s) and Interval Markov Chains ({\imc}s) by taking into
  account imprecision in the transition probability values:
  transitions in {\pimc}s are labeled with parametric intervals of
  probabilities.  In this work, we study the difference between
  {\pimc}s and other Markov Chain abstractions models and investigate
  the two usual semantics for {\imc}s: once-and-for-all and
  at-every-step. In particular, we prove that both semantics agree on
  the maximal/minimal reachability probabilities of a given \imc.  We
  then investigate solutions to several parameter synthesis problems
  in the context of {\pimc}s -- consistency, qualitative reachability
  and quantitative reachability -- that rely on constraint
  encodings. Finally, we propose a prototype implementation of our
  constraint encodings with promising results.
     
\end{abstract}

\section{Introduction}

Discrete time Markov chains ({\dtmc}s for short) are a standard
probabilistic modeling formalism that has been extensively used 
in the litterature 
to reason about software~\cite{whittaker1994markov} and real-life
systems~\cite{Husmeier2010}. However, when modeling real-life systems, the exact
value of transition probabilities may not be known precisely. Several
formalisms abstracting {\dtmc}s have therefore been
developed. Parametric Markov chains~\cite{Alur93} ({\pmc}s for short)
extend {\dtmc}s by allowing parameters to appear in transition
probabilities. In this formalism, parameters are variables and
transition probabilities may be expressed as polynomials over these
variables. A given {\pmc} therefore represents a potentially
infinite set of {\dtmc}s, obtained by replacing each parameter by a
given value. {\pmc}s are particularly useful to represent systems
where dependencies between transition probabilities are
required. Indeed, a given parameter may appear in several dinstinct
transition probabilities, therefore requiring that the same value is
given to all its occurences. Interval Markov chains~\cite{JonssonL91} ({\imc}
for short)
extend {\dtmc}s by allowing precise
transition probabilities to be replaced by intervals, but cannot represent dependencies between distinct transitions. {\imc}s have mainly
been studied with two distinct semantics interpretation. Under the
{\em once-and-for-all} semantics, a given {\imc} represents a
potentially infinite number of {\dtmc}s where transition probabilities
are chosen inside the specified intervals while keeping the same
underlying graph structure. The {\em at-every-step} semantics, which
was the original semantics given to {\imc}s in~\cite{JonssonL91}, does not
require {\dtmc}s to preserve the underlying graph structure of the
original {\imc} but instead allows an ``unfolding'' of the original
graph structure where different probability values may be chosen
(inside the specified interval) at each occurence of the given
transition.

Model-checking algorithms and tools have been developed 
in the context of
{\pmc}s~\cite{Prophesy,DBLP:conf/cav/HahnHWZ10,DBLP:conf/cav/KwiatkowskaNP11}
and {\imc}s with the once-and-for-all
semantics~\cite{Chakraborty2015,benedikt2013ltl}. State of the art tools~\cite{Prophesy} for {\pmc}
verification compute a rational function on the parameters that
characterizes the probability of satisfying a given property, and then
use external tools such as SMT solving~\cite{Prophesy} for computing the
satisfying parameter values. For these methods to be viable in
practice, the number of parameters used is quite limited. On the other
hand, the model-checking procedure for {\imc}s presented
in~\cite{benedikt2013ltl} is adapted from machine learning and builds
successive refinements of the original {\imc}s that optimize the
probability of satisfying the given property. This algorithm
converges, but not necessarilly to a global optimum. It is worth noticing that
existing model checking procedures for {\pmc}s and {\imc}s
strongly rely on their underlying graph structure. As a consequence,
to the best of our knowledge, no solutions for model-checking {\imc}s
with the at-every-step semantics have been proposed yet.

In this paper, we focus on Parametric interval Markov
chains~\cite{DelahayeLP16} ({\pimc}s for short), that generalize both
{\imc}s and {\pmc}s by allowing parameters to appear in the endpoints
of the intervals specifying transition probabilities, and we provide four main contributions. 

First, we formally compare abstraction formalisms for
{\dtmc}s in terms of succinctness: we show in particular that {\pimc}s
are {\em strictly more succinct} than both {\pmc}s and {\pimc}s when
equipped with the right semantics. In other words, everything that can
be expressed using {\pmc}s or {\imc}s can also be expressed using
{\pimc}s while the reverse does not hold. Second, we prove that the
once-and-for-all and the at-every-step semantics are equivalent w.r.t.
rechability properties, both in the {\imc} and in the {\pimc}
settings. Notably, this result gives theoretical backing to the
generalization of existing works on the verification of {\imc}s to the
at-every-step semantics.  Third, we study the parametric verification of
fundamental properties at the {\pimc} level: consistency, qualitative
reachability, and quantitative reachability. Given the expressivity of
the {\pimc} formalism, the risk of producing a {\pimc} specification
that is incoherent and therefore does not model any concrete {\dtmc}
is high. We therefore propose constraint encodings for deciding
whether a given {\pimc} is consistent and, if so, synthesizing
parameter values ensuring consistency. We then extend these encodings
to qualitative reachability, \ie ensuring that given state
labels are reachable in {\em all} (resp. {\em none}) of the {\dtmc}s
modeled by a given {\pimc}. Finally, we focus on the quantitative
reachability problem, \ie synthesizing parameter values such
that the probability of reaching given state labels satisfies fixed
bounds in {\em at least one} (resp. {\em all}) {\dtmc}s modeled by a
given {\pimc}. While consistency and qualitative reachability for
{\pimc}s have already been studied in~\cite{DelahayeLP16}, the
constraint encodings we propose in this paper are significantly
smaller (linear instead of exponential). To the best of our knowledge,
our results provide the first solution to the quantitative reachability problem for {\pimc}s.
Our last contribution is the implementation of all our verification algorithms in a prototype tool
that generates the required constraint encodings and can be plugged to
any SMT solver for their resolution.
Due to space limitation, the proofs of our results are given in Appendix.

\section{Background}\label{sec:background}

In this section we introduce notions and notations that will be used
throughout the paper. Given a finite set of variables $X = \{x_1, \ldots, x_k\}$, we write
$D_x$ for the domain of the variable $x \in X$ and $D_X$ for the set
of domains associated to the variables in $X$. A valuation
$v$ over $X$ is a set $v = \{(x,d) | x \in X, d \in D_x\}$ of
elementary valuations $(x,d)$ where for each $x \in X$ there exists a
unique pair of the form $(x, d)$ in $v$. When clear from the context,
we write $v(x) = d$ for the value given to variable $x$ according to
valuation $v$. A rational function $f$ over $X$ is a division of two
(multivariate) polynomials $g_1$ and $g_2$ over $X$ with rational
coefficients, \ie $f = g_1 / g_2$. We write ${\Qset}$ the set of 
 rational numbers and ${\Qset}_X$ the set of 
rational functions over $X$.  The evaluation $v(g)$ of a polynomial
$g$ under the valuation $v$ replaces each variable $x \in X$ by its value $v(x)$.

An {\em atomic constraint} over $X$ is a Boolean expression of the
form $f(X) \bowtie g(X)$, with ${\bowtie} \in \{\le, \ge, <, >, =\}$ and
$f$ and $g$ two functions over variables in $X$ and constants. A
constraint is {\em linear} if the functions $f$ and $g$ are
linear. 
A {\em constraint} over $X$ is a
Boolean combination of atomic constraints over $X$.

Given a finite set of states $S$, we write $\Dist(S)$ for the set of
probability distributions over $S$, \ie the set of functions $\mu : S
\to [0,1]$ such that $\sum_{s\in S}\mu(s) = 1$. We write $\Inter$
for the set containing all the interval subsets of $[0,1]$.  In the
following, we consider a universal set of symbols $A$ that we use for
labelling the states of our structures. We call these symbols {\em
  atomic propositions}.  We will use Latin alphabet in state context
and Greek alphabet in atomic proposition context.

\custompar{Constraints.}
Constraints are first order logic predicates used to model and solve combinatorial problems \cite{Rossi2006HCP}. 
A problem is described with a list of variables, each in a given domain of possible values, together
with a list of constraints over these variables. Such problems are then sent to solvers which decide whether the problem is
satisfiable, \ie if there exists a valuation of the variables
satisfying all the constraints, and in this case computes a solution. 
Checking satisfiability of constraint problems is difficult in general, as the space of all possible valuations has a size exponential in the number of variables.
  
  Formally, a Constraint Satisfaction Problem
(\csp) is a tuple $\Omega = (X, D, C)$ where $X$ is a finite set of
variables, $D = D_X$ is the set of all the domains associated to the
variables from $X$, and $C$ is a set of constraints over $X$.  We say
that a valuation over $X$ satisfies $\Omega$ if and only if it
satisfies all the constraints in $C$.  We write $v(C)$ for the
satisfaction result of the valuation of the constraints $C$ according
to $v$ (\ie true or false).
%
%
In the following we call {\em {\csp} encoding} a scheme for formulating a given problem into a {\csp}.
The size of a {\csp} corresponds to the number of variables and atomic constraints appearing in the problem.
Note that, in constraint programming, having less variables or less constraints during the encoding does not necessarily imply
faster solving time of the problems.

\custompar{Discrete Time Markov Chains.}
A Discrete Time Markov Chain ({DTMC} or {\mc} for short) is a tuple
$\mathcal{M}$ $=$ $(S,$ $s_0,$ $p,$ $V)$, where $S$ is a finite
set of states containing the initial state $s_0$, 
$V : S \to 2^A$ is a labelling function, and
$p : S \rightarrow \Dist(S)$ is a probabilistic transition
function. We write
{\mcSet} for the set containing all the discrete time Markov chains.

A Markov Chain can be seen as a directed graph
where the nodes correspond to the states of the {\mc} and
the edges are labelled with the probabilities given by the transition function of the {\mc}.
In this representation, a missing transition between two states 
represents a transition probability of zero.
As usual, given a {\mc}~$\mathcal{M}$, we call a {\em path} of
$\mathcal{M}$ a sequence
of states obtained from executing $\mathcal{M}$, \ie a sequence
$\omega = s_1, s_2,\ldots $ s.t. the probability of taking the transition from $s_i$ to $s_{i+1}$ is strictly positive,   $p(s_i)(s_i+1) >0$, for all $i$.
A path $\omega$ is finite iff it belongs to $S^*$, \ie it
represents a finite sequence of transitions from $\mathcal{M}$.

\begin{example}\label{ex:mc}
Figure \ref{fig:example_mc} illustrates the Markov chain 
$\mathcal{M}_1 = (S, s_0, p, V) \in \mcSet$ where
the set of states $S$ is given by $\{s_0,s_1,s_2,s_3,s_4\}$, 
the atomic proposition are restricted to $\{\alpha, \beta\}$, 
the initial state is $s_0$, and the labelling 
function $V$ corresponds to $\{(s_0,\emptyset), (s_1,\alpha), (s_2,\beta), (s_3,\{\alpha, \beta\}), (s_4,\alpha)\}$.
The sequences of states $(s_0,s_1,s_2)$, $(s_0,s_2)$, and $(s_0,s_2,s_2,s_2)$, 
are three (finite) paths from the initial state $s_0$ to the state $s_2$.
\end{example}

\custompar{Reachability.}
A Markov chain $\mathcal{M}$ defines a unique probability measure
$\mathbb{P}^{\mathcal{M}}$ over the paths from
$\mathcal{M}$. According to this measure, the
probability of a finite path $\omega = s_0, s_1, \ldots, s_n$ in
$\mathcal{M}$ is the product of the probabilities of the transitions
executed along this path, \ie $\mathbb{P}^{\mathcal{M}}(\omega) =
p(s_0)(s_1) \cdot p(s_1)(s_2)\cdot \ldots \cdot p(s_{n-1})(s_n)$. This
distribution naturally extends to infinite paths (see
\cite{Baier2008PMC}) and to sequences of states over $S$
that are not paths of $\mathcal{M}$ by giving them a zero probability.

Given a \mc\ $\mathcal{M}$, the overall probability of reaching a
given state $s$ from the initial state $s_0$ is called the {\em
  reachability probability} and written
$\mathbb{P}^{\mathcal{M}}_{s_0}(\ltlExists s)$ or
$\mathbb{P}^{\mathcal{M}}(\ltlExists s)$ when clear from the
context. This probability is computed as the sum of the probabilities
of all finite paths starting in the initial state and reaching this
state for the first time. Formally, let $\mathsf{reach}_{s_0}(s) =
\{\omega \in S^{*} \st \omega = s_0, \ldots s_n \mbox{ with } s_n = s
\mbox{ and } s_i \ne s \ \forall 0 \le i < n\}$ be the set of such
paths. We then define $\mathbb{P}^{\mathcal{M}}(\ltlExists s) =
\sum_{\omega \in \mathsf{reach}_{s_0}(s)} \mathbb{P}^{\mathcal{M}}(\omega)$
if $s \ne s_0$ and $1$ otherwise. This notation naturally extends to
the reachability probability of a state $s$ from a state $t$ that is
not $s_0$, written $\mathbb{P}^{\mathcal{M}}_{t}(\ltlExists s)$
and to the
probability of reaching a label $\alpha \subseteq A$ written
$\mathbb{P}^{\mathcal{M}}_{s_0}(\ltlExists \alpha)$.
In the following, we say that a state $s$ (resp. a label $\alpha \subseteq A$) is reachable in $\mathcal{M}$
iff the reachability probability of this state (resp. label)
from the initial state is strictly positive.

\begin{example}[Example \ref{ex:mc} continued]
In Figure \ref{fig:example_mc} the probability of the path $(s_0,$
$s_2,$ $s_1,$ $s_1,$ $s_3)$ is $0.3 \cdot 0.5 \cdot 0.5 \cdot 0.5 =
0.0375$ and the probability of reaching the state $s_1$ is
$\Proba^{\mathcal{M}_1}(\ltlExists s_1) = p(s_0)(s_1) +
\Sigma_{i=0}^{+\infty}{p(s_0)(s_2){\cdot}p(s_2)(s_2)^i{\cdot}p(s_2)(s_1)}
= p(s_0)(s_1) + p(s_0)(s_2){\cdot}p(s_2)(s_1){\cdot}(1/(1-p(s_2)(s_2)))
= 1$.  Furthermore, the probability of reaching $\beta$ corresponds to
the probability of reaching the state $s_2$.
\end{example}

\begin{figure*}[t]
\mbox{
\begin{minipage}[t]{.31\textwidth}
\begin{center}
{\tiny
\begin{tikzpicture}[scale=1.5]
	\node[vertex] (n0) at (0,0.5) {$s_0$};
	\node[vertex, label={[label distance=-0.03cm]-45 :$\alpha$}] (n1) at (0.9,1)   {$s_1$};
	\node[vertex, label={[label distance=-0.10cm]+45 :$\beta$}] (n2) at (0.9,0)   {$s_2$};
	\node[vertex, label={[label distance=-0.10cm]-135:$\alpha,\beta$}] (n3) at (1.8,1)   {$s_3$};
	\node[vertex, label={[label distance=-0.10cm]+135:$\alpha$}] (n4) at (1.8,0)   {$s_4$};

	\draw[edge] (n0) to node[above left] {$0.7$} (n1);
	\draw[edge] (n0) to node[below left] {$0.3$} (n2);

	\draw[edge] (n1) to[out=45,in=45+90,min distance=4mm] node[above] {$0.5$} (n1);
	\draw[edge] (n1) to node[above] {$0.5$} (n3);
	
	\draw[edge] (n2) to node[right] {$0.5$} (n1);
	\draw[edge] (n2) to[out=-45,in=-45-90,min distance=4mm] node[below] {$0.5$} (n2);
	
	\draw[edge] (n3) to[out=45,in=45+90,min distance=4mm] node[above] {$1$} (n3);
	
	\draw[edge] (n4) to[out=-45,in=-45-90,min distance=4mm] node[below] {$1$} (n4);
\end{tikzpicture}}
\caption{{\mc} $\mathcal{M}_1$}\label{fig:example_mc}%
\end{center}
\end{minipage}}
%
\mbox{
\begin{minipage}[t]{.3\textwidth}
\begin{center}
{\tiny
\begin{tikzpicture}[scale=1.5]
	\node[vertex] (n0) at (0,0.5) {$s_0$};
	\node[vertex, label={[label distance=-0.03cm]-45 :$\alpha$}] (n1) at (0.9,1)   {$s_1$};
	\node[vertex, label={[label distance=-0.10cm]+45 :$\beta$}] (n2) at (0.9,0)   {$s_2$};
	\node[vertex, label={[label distance=-0.10cm]-135:$\alpha,\beta$}] (n3) at (1.8,1)   {$s_3$};
	\node[vertex, label={[label distance=-0.10cm]+135:$\alpha$}] (n4) at (1.8,0)   {$s_4$};

	\draw[edge] (n0) to node[above left] {$0.7$} (n1);
	\draw[edge] (n0) to node[below left] {$0.3$} (n2);

	\draw[edge] (n1) to[out=45,in=45+90,min distance=4mm] node[above] {$1-p$} (n1);
	\draw[edge] (n1) to node[above] {$p$} (n3);
	
	\draw[edge] (n2) to node[right] {$p$} (n1);
	\draw[edge] (n2) to[out=-45,in=-45-90,min distance=4mm] node[below] {$1-p$} (n2);
	
	\draw[edge] (n3) to[out=45,in=45+90,min distance=4mm] node[above] {$1$} (n3);
	
	\draw[edge] (n4) to[out=-45,in=-45-90,min distance=4mm] node[below] {$1$} (n4);
\end{tikzpicture}}
\caption{\pmc\ $\mathcal{I}^\prime$}\label{fig:example_pmc}%
\end{center}
\end{minipage}}
\mbox{
\begin{minipage}[t]{.35\textwidth}
\begin{center}
{\tiny
\hspace*{-0.1cm}
\begin{tikzpicture}[scale=1.5]
	\node[vertex] (n0) at (0,0.5) {$s_0$};
	\node[vertex, label={[label distance=-0.03cm]-45 :$\alpha$}] (n1) at (0.9,1)   {$s_1$};
	\node[vertex, label={[label distance=-0.10cm]+45 :$\beta$}] (n2) at (0.9,0)   {$s_2$};
	\node[vertex, label={[label distance=-0.10cm]-135:$\alpha, \beta$}] (n3) at (2.1,1)   {$s_3$};
	\node[vertex, label={[label distance=-0.10cm]+135:$\alpha$}] (n4) at (2.1,0)   {$s_4$};

	\draw[edge] (n0) to node[above left] {$[0,1]$} (n1);
	\draw[edge] (n0) to node[below left] {$[0,1]$} (n2);

	\draw[edge] (n1) to[out=45,in=45+90,min distance=4mm] node[above] {$[0.5,1]$} (n1);
	\draw[edge] (n1) to node[above] {$[0.3,0.5]$} (n3);
	
	\draw[edge] (n2) to node[right] {$[0,0.6]$} (n1);
	\draw[edge] (n2) to[out=-45,in=-45-90,min distance=4mm] node[below] {$[0.2,0.6]$} (n2);
	\draw[edge] (n2) to node[below] {$[0,0.5]$} (n4);	
	
	\draw[edge] (n3) to[out=45,in=45+90,min distance=4mm] node[above] {$1$} (n3);
	
	\draw[edge] (n4) to node[right] {$[0,0.5]$} (n3);
	\draw[edge] (n4) to[out=-45,in=-45-90,min distance=4mm] node[below] {$[0.5,0.6]$} (n4);
\end{tikzpicture}}
\caption{{\imc} $\mathcal{I}$}\label{fig:example_imc}
\end{center}
\end{minipage}}
\vspace*{-0.4cm}
\end{figure*}

\section{Markov Chains Abstractions}
\label{sec:abstraction-models}

Modelling an application as a Markov Chain requires knowing the exact
probability for each possible transition of the system.  However, this
can be difficult to compute or to measure in the case of a real-life
application (\eg precision errors, limited knowledge).  In this
section, we start with a generic definition of Markov chain
abstraction models. Then we recall three abstraction models from the
literature, respectively {\pmc}, {\imc}, and {\pimc}, and finally we
present a comparison of these existing models in terms of succinctness.

\begin{definition}[Markov chain Abstraction Model]\label{def:abstract_model}
	A Markov chain abstraction model (an abstraction model for short) is a pair $(\aminstance, \satisfaction)$ 
    where $\aminstance$ is a nonempty set and $\satisfaction$ is a relation between ${\mcSet}$ and $\aminstance$. 
    Let $\mathcal{P}$ be in $\aminstance$ and $\mathcal{M}$ be in {\mcSet}
    we say that $\mathcal{M}$ implements $\mathcal{P}$ 
    iff $(\mathcal{M}, \mathcal{P})$ belongs to $\satisfaction$ (\ie $\mathcal{M} \satisfaction \mathcal{P}$).
    When the context is clear, we do not mention the satisfaction relation $\satisfaction$ 
    and only use $\aminstance$ to refer to the abstraction model $(\aminstance, \satisfaction)$.
\end{definition}

A {\em Markov chain Abstraction Model} is a specification theory for
{\mc}s. It consists in a set of abstract objects, called {\em
  specifications}, each of which representing a (potentially infinite)
set of {\mc}s -- {\em implementations} -- together with a satisfaction
relation defining the link between implementations and specifications.
As an example, consider the powerset of {\mcSet} (\ie the set
containing all the possible sets of Markov chains). Clearly,
$(2^{\mcSet}, \in)$ is a Markov chain abstraction model, which we call
the {\em canonical abstraction model}.  This abstraction model has the
advantage of representing all the possible sets of Markov chains but
it also has the disadvantage that some Markov chain abstractions are
only representable by an infinite extension representation.  Indeed,
recall that there exists subsets of $[0,1] \subseteq \Rset$ which
cannot be represented in a finite space (e.g., the Cantor set
\cite{Cantor1883}).
We now present existing {\mc} abstraction models from the literature.

\subsection{Existing MC Abstraction Models}

\custompar{Parametric Markov Chain}
is a {\mc} abstraction model 
from \cite{Alur93} 
where a transition can be annotated by a
rational function over {\em parameters}. 
We write $\pmcSet$ for the set containing
all the parametric Markov chains.

\begin{definition}[Parametric Markov Chain]\label{def:pmc}
A Parametric Markov Chain ({\pmc} for short) 
is a tuple $\mathcal{I} = (S,s_0,P,V,Y)$
where $S$, $s_0$, and $V$ are defined as for \mc{s}, 
$Y$ is a set of variables (parameters),
and $P: S \times S \to \Qset_Y$  associates with each potential transition 
a parameterized probability.
\end{definition}

Let $\mathcal{M} = (S,s_0,p,V)$ be a $\mc$ and $\mathcal{I} =
(S,s_0,P,V,Y)$ be a $\pmc$.  The satisfaction relation
$\satisfactionPmc$ between $\mcSet$ and $\pmcSet$ is defined by
$\mathcal{M} \satisfactionPmc \mathcal{I}$ iff there exists a
valuation $v$ of $Y$ s.t. $p(s)(s^\prime)$ equals $v(P(s,s^\prime))$
for all $s,s^\prime$ in $S$.  
\begin{example}
	Figure~\ref{fig:example_pmc} shows a {\pmc} $\mathcal{I}^\prime = (S,s_0,P,V,Y)$
    where $S$, $s_0$, and $V$ are similar to the same entities in the {\mc} $\mathcal{M}$
    from Figure~\ref{fig:example_mc}, the set of variable $Y$ contains only one variable $p$,
    and the parametric transitions in $P$ are given by the edge labelling
    (\eg $P(s_0,s_1) = 0.7$, $P(s_1,s_3) = p$, and $P(s_2,s_2) = 1 - p$).
    Note that the {\pmc} $\mathcal{I}^\prime$ is a specification
    containing the {\mc} $\mathcal{M}$ from Figure~\ref{fig:example_mc}.
\end{example}

\custompar{Interval Markov Chains} 
extend {\mc}s by allowing to label transitions with intervals of possible
probabilities instead of precise probabilities.
We write
$\imcSet$ for the set containing all the interval Markov chains.

\begin{definition}[Interval Markov Chain \cite{JonssonL91}]\label{def:imc}
	An Interval Markov Chain ({\imc} for short) is a tuple $\mathcal{I} = (S,s_0,P,V)$, 
	where $S$, $s_0$, and $V$ are defined as for \mc{s}, 
	and $P : S \times S \to \Inter$ associates 
	with each potential transition an interval of probabilities.
\end{definition}

\begin{example}\label{ex:imc}
Figure \ref{fig:example_imc} illustrates \imc\ $\mathcal{I} = (S, s_0, P, V)$ where
$S$, $s_0$, and $V$ are similar to the \mc\ given in Figure \ref{fig:example_mc}.
By observing the edge labelling
we see that $P(s_0,s_1)=[0,1]$, $P(s_1,s_1)=[0.5,1]$, and $P(s_3,s_3)=[1, 1]$.
On the other hand, the intervals of probability for missing transitions are reduced to $[0,0]$, e.g., $P(s_0,s_0)=[0,0]$, $P(s_0,s_3)=[0,0]$, $P(s_1,s_4)=[0,0]$.
\end{example}

In the literature, {\imc}s have been mainly used with two distinct
semantics: {\em at-every-step} and {\em once-and-for-all}. Both
semantics are associated with distinct satisfaction relations which we
now introduce.

The {\em once-and-for-all} {\imc} semantics (\cite{Prophesy,tulip,puggelli13}) is alike to the semantics for \pmc, as
introduced above. The associated satisfaction relation
$\satisfactionImcOnce$ is defined as follows: A {\mc} $\mathcal{M} =
(T, t_0, p, V^M)$ satisfies an {\imc} $\mathcal{I} =
(S,s_0,P,V^I)$ iff $(T,t_0,V^M) = (S,s_0,V^I)$ and for all
reachable state $s$ and all state $s' \in S$, $p(s)(s') \in
P(s,s')$. In this sense, we say that {\mc} implementations using the
once-and-for-all semantics need to have the same structure as the
{\imc} specification.

On the other hand, the {\em at-every-step} {\imc} semantics, first introduced
in~\cite{JonssonL91}, operates as a simulation relation based on the
transition probabilities and state labels, and therefore allows {\mc}
implementations to have a different structure than the {\imc}
specification. The associated satisfaction relation $\satisfactionImc$
is defined as follows: A {\mc} $\mathcal{M}$ $=$ $(T,
t_0,$ $p,$ $V^M)$ satisfies an {\imc} $\mathcal{I} = (S,s_0,P,V^I)$ iff
there exists a relation $\mathcal{R} \subseteq T \times S$ such that
$(t_0, s) \in \mathcal{R}$ and 
whenever $(t,s) \in \mathcal{R}$, we have \begin{enumerate*}
\item the labels of $s$ and $t$ correspond: $V^M(t) = V^I(s)$,
\item there exists a correspondence function $\delta: T \to (S \to [0, 1])$ s.t.
  \begin{enumerate*}
  \item $\forall t^\prime \in T$ if $p(t)(t^\prime) > 0$ then $\delta(t^\prime)$ is a distribution on $S$
  \item $\forall s^\prime \in S:
    (\Sigma_{t^\prime \in T} p(t)(t^\prime) \cdot \delta(t^\prime)(s^\prime)) \in P(s,s^\prime)$, and 
  \item $\forall (t^\prime,s^\prime) \in T \times S$, if $\delta(t^\prime)(s^\prime) > 0$, 
    then $(t^\prime, s^\prime) \in \mathcal{R}$.
    
  \end{enumerate*}
\end{enumerate*}
By construction, it is clear that $\satisfactionImc$ is more general
than $\satisfactionImcOnce$, \ie that whenever $\mathcal{M}
\satisfactionImcOnce \mathcal{I}$, we also have $\mathcal{M}
\satisfactionImc \mathcal{I}$. The reverse is obviously not true in
general, even when the underlying graphs of $\mathcal{M}$ and
$\mathcal{I}$ are isomorphic (see Appendix~\ref{ap:compare_imcs_satisfaction_relations} for details).

\begin{figure*}[t]
\mbox{
\begin{minipage}[t]{.48\textwidth}
\begin{center}
{\tiny
\begin{tikzpicture}[scale=1.5]
	\node[vertex] (n0) at  (0,0.5) {$t_0$};
	\node[vertex, label={[label distance=-0.03cm]-105 :$\alpha$}] (n1) at  (0.9,1)   {$t_1$};
	\node[vertex, label={[label distance=-0.05cm]0    :$\beta$}] (n2) at  (0.9,0)   {$t_2$};
	\node[vertex, label={[label distance=-0.05cm]0    :$\alpha,\beta$}] (n3) at  (2,1.3)   {$t_3$};
	\node[vertex, label={[label distance=-0.05cm]0    :$\alpha,\beta$}, text height=1.3ex,text width=0.6em] (n3p) at (2,0.7)     {$t_{3^\prime}$};

	\draw[edge] (n0) to node[above left] {$0.7$} (n1);
	\draw[edge] (n0) to node[below left] {$0.3$} (n2);

	\draw[edge] (n1) to[out=45,in=45+90,min distance=4mm] node[above] {$0.5$} (n1);
	\draw[edge] (n1) to node[above] {$0.3$} (n3);
	\draw[edge] (n1) to node[below] {$0.2$} (n3p);
	
	\draw[edge] (n2) to node[right] {$0.8$} (n1);
	\draw[edge] (n2) to[out=-45,in=-45-90,min distance=4mm] node[below] {$0.2$} (n2);
	
	\draw[edge] (n3) to[out=45+90,in=45,min distance=4mm] node[above] {$0.2$} (n3);
	\draw[edge] (n3) to node[right] {$0.8$} (n3p);	

	\draw[edge] (n3p) to[out=-45-90,in=-45,min distance=4mm] node[below] {$1$} (n3p);

\end{tikzpicture}}
\caption{\mc\ $\mathcal{M}_2$ satisfying the \imc\ $\mathcal{I}$ from Figure \ref{fig:example_imc} with a different structure}\label{fig:example_mc_not_iso}
\end{center}
\end{minipage}}
\hspace{.005\textwidth}
\mbox{
\begin{minipage}[t]{.46\textwidth}
\begin{center}
{\tiny
\begin{tikzpicture}[scale=1.4]
	\node[vertex] (n0) at (0,0.5) {$s_0$};
	\node[vertex, label={[label distance=-0.03cm]-45 :$\alpha$}] (n1) at (0.9,1)   {$s_1$};
	\node[vertex, label={[label distance=-0.10cm]+45 :$\beta$}] (n2) at (0.9,0)   {$s_2$};
	\node[vertex, label={[label distance=-0.10cm]-135:$\alpha, \beta$}] (n3) at (2.1,1)   {$s_3$};
	\node[vertex, label={[label distance=-0.10cm]+135:$\alpha$}] (n4) at (2.1,0)   {$s_4$};

	\draw[edge] (n0) to node[above left] {$[0,1]$} (n1);
	\draw[edge] (n0) to node[below left] {$[0,1]$} (n2);

	\draw[edge] (n1) to[out=45,in=45+90,min distance=4mm] node[above] {$[q,1]$} (n1);
	\draw[edge] (n1) to node[above] {$[0.3,q]$} (n3);
	
	\draw[edge] (n2) to node[right] {$[0,p]$} (n1);
	\draw[edge] (n2) to[out=-45,in=-45-90,min distance=4mm] node[below] {$[0.2,p]$} (n2);
	\draw[edge] (n2) to node[below] {$[0,0.5]$} (n4);	
	
	\draw[edge] (n3) to[out=45,in=-45,min distance=4mm] node[right] {$1$} (n3);
	
	\draw[edge] (n4) to node[right] {$[0,0.5]$} (n3);
	\draw[edge] (n4) to[out=-45,in=45,min distance=4mm] node[right] {$[0.5,p]$} (n4);
\end{tikzpicture}}
\caption{{\pimc} $\mathcal{P}$}\label{fig:example_pimc}
\end{center}
\end{minipage}}
\vspace*{-0.4cm}
\end{figure*}

\begin{example}[Example \ref{ex:imc} continued]\label{ex:mcs_satify_imc}
Consider the {\mc} $\mathcal{M}_1$ with state space $S$ from Figure \ref{fig:example_mc} and 
the {\mc} $\mathcal{M}_2$ with state space $T$ from Figure \ref{fig:example_mc_not_iso}. They both 
satisfy the \imc\ $\mathcal{I}$ with state space $S$ given in Figure \ref{fig:example_imc}.
Furthermore, $\mathcal{M}_1$ satisfies $\mathcal{I}$ with the same structure.
On the other hand, for the {\mc} $\mathcal{M}_2$ given in Figure \ref{fig:example_mc_not_iso}, 
the state $s_3$ from $\mathcal{I}$ has been ``split'' into two states $t_3$ and $t_{3^\prime}$ in $\mathcal{M}_2$
and the state $t_1$ from $\mathcal{M}_2$ ``aggregates'' states $s_1$ and $s_4$ in $\mathcal{I}$.
The relation $\mathcal{R} \subseteq T \times S$ containing the pairs 
$(t_0,s_0)$, $(t_1,s_1)$, $(t_1,s_4)$, $(t_2,s_2)$, $(t_3,s_3)$, and $(t_{3^\prime}, s_3)$ 
is a satisfaction relation between $\mathcal{M}_2$ and $\mathcal{I}$.
\end{example}

\custompar{Parametric Interval Markov Chains},
as introduced in \cite{DelahayeLP16}, abstract {\imc}s by allowing (combinations of)
parameters to be used as interval endpoints in {\imc}s. Under a given
parameter valuation the {\pimc} yields an {\imc} as introduced above.
{\pimc}s therefore allow the representation, in a compact way and with a
finite structure, of a potentially infinite number of {\imc}s. Note that
one parameter can appear in several transitions at once, requiring the
associated transition probabilities to depend on one another.
Let $Y$ be a finite set of
parameters and $v$ be a valuation over $Y$.  By combining notations
used for $\imc$s and $\pmc$s the set $\Inter(\Qset_Y)$ contains all
parametrized intervals over $[0,1]$, and for all $I
= [f_1, f_2] \in \Inter(\Qset_Y)$, $v(I)$ denotes the interval
$[v(f_1), v(f_2)]$ if $0 \le v(f_1) \leq v(f_2) \le 1$ and the empty set
otherwise\footnote{Indeed, when $0 \le v(f_1) \leq v(f_2) \le 1$ is not respected, the interval is inconsistent and therefore empty.}.
We write $\pimcSet$ for the set containing all the parametric interval Markov chains.


\begin{definition}[Parametric Interval Markov Chain \cite{DelahayeLP16}]\label{def:pimc}
	A Parametric Interval Markov Chain ({\pimc} for short) is a tuple $\mathcal{P} = (S,s_0,P,V,Y)$, 
	where $S$, $s_0$, $V$ and $Y$ are defined as for {\pmc}s, 
	and 
	$P : S \times S \to \Inter(\Qset_Y)$ 
    associates with each potential transition a (parametric) interval.
\end{definition}

In \cite{DelahayeLP16} the authors introduced {\pimc}s where parametric interval
endpoints are limited to linear combination of parameters.  In this paper we extend the {\pimc} model by allowing rational
functions over parameters as endpoints of parametric intervals.  Given
a \pimc\ $\mathcal{P} =(S,s_0,P,V,Y)$ and a valuation $v$, we write
$v(\mathcal{P})$ for the \imc\ $(S,s_0,P_v,V)$ obtained by replacing
the transition function $P$ from $\mathcal{P}$ with the function $P_v
: S \times S \to \Inter$ defined by $P_v(s,s^\prime) =
v(P(s,s^\prime))$ for all $s,s^\prime \in S$.
The \imc\ $v(\mathcal{P})$ is called an {\em instance} of
\pimc\ $\mathcal{P}$.  Finally, depending on the semantics chosen for
       {\imc}s, two satisfaction relations can be defined between
       {\mc}s and {\pimc}s. They are written $\satisfactionPimc$ and
       $\satisfactionPimcOnce$ and defined as follows: $\mathcal{M}
       \satisfactionPimc \mathcal{P}$ (resp. $\satisfactionPimcOnce$)
       iff there exists an \imc\ $\mathcal{I}$ instance of
       $\mathcal{P}$ s.t. $\mathcal{M} \satisfactionImc
       \mathcal{I}$ (resp. $\satisfactionImcOnce$).

\begin{example}
Consider the {\pimc} $\mathcal{P} = (S,$0$,P,V,Y)$ given in Figure \ref{fig:example_pimc}.
The set of states $S$ and the labelling function
are the same as in the ${\mc}$ and the ${\imc}$  presented in Figures~\ref{fig:example_mc}~and~\ref{fig:example_imc} respectively.
The set of parameters $Y$ has two elements $p$ and $q$.
Finally, the parametric intervals from the transition function $P$ 
are given by the edge labelling
(\eg $P(s_1,s_3)=[0.3, q]$, $P(s_2,s_4)=[0,0.5]$, and $P(s_3,s_3)=[1,1]$).
Note that the {\imc} $\mathcal{I}$ from Figure \ref{fig:example_imc} 
is an instance of $\mathcal{P}$
(by assigning the value $0.6$ to the parameter $p$ and $0.5$ to  $q$).
Furthermore, as said in Example \ref{ex:mcs_satify_imc}, 
the Markov Chains $\mathcal{M}_1$ and $\mathcal{M}_2$ (from Figures \ref{fig:example_mc} and \ref{fig:example_mc_not_iso} respectively)
satisfy $\mathcal{I}$, therefore $\mathcal{M}_1$ and $\mathcal{M}_2$ satisfy $\mathcal{P}$.
\end{example}

In the following, we consider that the size of a {\pmc}, {\imc}, or {\pimc}
corresponds to its number of states plus its number of transitions not reduced to $0$, $[0,0]$ or $\emptyset$.
We will also often need to consider the predecessors (\Pred), 
and the successors (\Succ) of some given states.
Given a \pimc\ with a set of states $S$, a state $s$ in $S$, and a subset $S^\prime$ of $S$, we write:

\smallskip
\begin{minipage}{\textwidth}
  \hspace{-.8cm}
  \begin{minipage}{.58\textwidth}
    \begin{itemize}
	\item { $\Pred(s) = \{s^\prime \in S \mid P(s^\prime, s)
          \notin \{\emptyset, [0, 0]\}\}$}
	\item { $\Succ(s) = \{s^\prime \in S \mid P(s,
          s^\prime) \notin \{\emptyset, [0,
            0] \}\}$}
    \end{itemize}
  \end{minipage}
  \begin{minipage}{.41\textwidth}
    \begin{itemize}
      \item { $\Pred(S^\prime) = \bigcup_{s^\prime \in
          S^\prime} \Pred(s^\prime)$}
      \item { $\Succ(S^\prime) = \bigcup_{s^\prime \in S^\prime}
        \Succ(s^\prime)$}
    \end{itemize}
  \end{minipage}  
\end{minipage}

\smallskip


\subsection{Abstraction Model Comparisons}

${\imc}$, ${\pmc}$, and ${\pimc}$ are three Markov chain Abstraction Models.
In order to compare their expressiveness and  compactness, we introduce
the comparison operators $\entailed$ and $\equiv$.
Let $(\aminstance_1, \models_1)$ and $(\aminstance_2,\models_2\nobreak)$ be two Markov chain abstraction models
containing respectively $\mathcal{L}_1$ and $\mathcal{L}_2$.
We say that $\mathcal{L}_1$ is entailed by $\mathcal{L}_2$, 
written $\mathcal{L}_1 \entailed \mathcal{L}_2$,
iff all the {\mc}s satisfying $\mathcal{L}_1$ satisfy $\mathcal{L}_2$
modulo bisimilarity.
(\ie 
$\forall \mathcal{M} \models_1 \mathcal{L}_1, \exists \mathcal{M}^\prime \models_2 \mathcal{L}_2$ s.t. $\mathcal{M}$ is bisimilar to $\mathcal{M}^\prime$).
We say that $\mathcal{L}_1$ is (semantically) equivalent to $\mathcal{L}_2$, 
written $\mathcal{L}_1 \equiv \mathcal{L}_2$,
iff $\mathcal{L}_1 \entailed \mathcal{L}_2$ and $\mathcal{L}_2 \entailed \mathcal{L}_1$.
Definition~\ref{def:succinctness} introduces succinctness based on the sizes of the abstractions.

\begin{definition}[Succinctness]\label{def:succinctness}
	Let $(\aminstance_1, \models_1)$ and $(\aminstance_2, \models_2)$ be two Markov chain abstraction models.
    $\aminstance_1$ is at least as succinct as $\aminstance_2$, 
    written $\aminstance_1 \leq \aminstance_2$, 
    iff there exists a polynomial $p$ such that for every 
    $\mathcal{L}_2 \in \aminstance_2$, 
    there exists $\mathcal{L}_1 \in \aminstance_1$ 
    s.t. $\mathcal{L}_1 \equiv \mathcal{L}_2$ and 
    $|\mathcal{L}_1| \leq p(|\mathcal{L}_2|)$.\footnote{
    $|\mathcal{L}_1|$ and $|\mathcal{L}_2|$ are the sizes of $\mathcal{L}_1$ and
    $\mathcal{L}_2$, respectively.}
    Thus, $\aminstance_1$ is strictly more succinct than $\aminstance_2$,
    written $\aminstance_1 < \aminstance_2$, iff
    $\aminstance_1 \leq \aminstance_2$
    and $\aminstance_2 \not\leq \aminstance_1$.
\end{definition}

We start with a comparison of the succinctness of the {\pmc} and {\imc}
abstractions. Since {\pmc}s allow the expression of dependencies
between the probabilities assigned to distinct transitions while
{\imc}s allow all transitions to be independant, it is clear that
there are {\pmc}s without any equivalent {\imc}s (regardless of the
{\imc} semantics used), therefore $(\imcSet,\satisfactionImcOnce) \not
\le \pmcSet$ and $(\imcSet,\satisfactionImc) \not \le \pmcSet$.  On
the other hand, {\imc}s imply that transition probabilities need to
satisfy linear inequalities in order to fit given intervals. However,
these types of constraints are not allowed in {\pmc}s. It is therefore
easy to exhibit {\imc}s that, regardless of the semantics considered,
do not have any equivalent {\pmc} specification. As a consequence, $
\pmcSet \not \le (\imcSet,\satisfactionImcOnce)$ and $ \pmcSet \not
\le (\imcSet,\satisfactionImc)$.


We now compare {\pmc}s and {\imc}s to {\pimc}s. Recall that the
{\pimcSet} model is a Markov chain abstraction model allowing to
declare parametric interval transitions, while the {\pmcSet} model
allows only parametric transitions (without intervals), and the
{\imcSet} model allows interval transitions without parameters.
Clearly, any {\pmc} and any {\imc} can be translated into a {\pimc}
with the right semantics (once-and-for-all for {\pmc}s and the chosen
{\imc} semantics for {\imc}s). This means that
$({\pimcSet},\satisfactionPimcOnce)$ is more succinct than {\pmcSet}
and {\pimcSet} is more succinct than {\imcSet} for both semantics.
Furthermore, since {\pmcSet} and {\imcSet} are not comparable due to
the above results, we have that the {\pimcSet}
abstraction model is strictly more succinct than the {\pmcSet}
abstraction model and than the {\imcSet} abstraction model with the
right semantics. Our comparison results are presented in Proposition~\ref{prop:sunccinctness_pimc_pmc_imc}.
Further explanations and examples
are given in Appendix~\ref{ap:model_comparison}.

\begin{proposition}\label{prop:sunccinctness_pimc_pmc_imc}
The Markov chain abstraction models can be ordered as follows w.r.t.
succinctness:
$({\pimcSet},\satisfactionPimcOnce) < (\pmcSet, \satisfactionPmc)$,
$({\pimcSet},\satisfactionPimcOnce) < (\imcSet, \satisfactionImcOnce)$ and
$({\pimcSet},\satisfactionPimc\nobreak) < (\imcSet, \satisfactionImc)$.
\end{proposition}

Note that $(\pmcSet, \satisfactionPmc) \le (\imcSet, \satisfactionImcOnce)$ could be achieved by adding unary constraints on the parameters of a {\pmc}, which is not allowed here. However, this would not have any impact on our other results.

\section{Qualitative Properties}\label{sec:qualitative-reachability}


As seen above, {\pimc}s are a succinct abstraction formalism for
{\mc}s. The aim of this section is to investigate qualitative
properties for {\pimc}s, \ie properties that can be evaluated at the
specification ({\pimc}) level, but that entail properties on its {\mc}
implementations.
{\pimc} specifications are very expressive as they allow the abstraction of
transition probabilities using both intervals and
parameters. Unfortunately, as it is the case for {\imc}s, this allows the
expression of incorrect specifications. In the {\imc} setting, this is
the case either when some intervals are ill-formed or when there is no
probability distribution matching the interval constraints of the
outgoing transitions of some reachable state. In this case, no {\mc}
implementation exists that satisfies the {\imc} specification. 
Deciding
whether an implementation 
that satisfies a given specification
exists  is
called the consistency problem. In the {\pimc} setting, the
consistency problem is made more complex because of the parameters
which can also induce inconsistencies in some cases. One could also be
interested in verifying whether there exists an implementation that
reaches some target states/labels, and if so, propose a parameter
valuation ensuring this property. Both the consistency and the
consistent reachability problems have already been investigated in the
{\imc} and {\pimc} setting~\cite{Delahaye15,DelahayeLP16}. In this section,
we briefly recall these problems and propose new solutions  based on CSP encodings. Our encodings are linear in the size
of the original {\pimc}s whereas the algorithms
  from~\cite{Delahaye15,DelahayeLP16} are exponential.

\subsection{Existential Consistency}

A {\pimc} $\mathcal{P}$ is existential consistent iff 
there exists a {\mc} $\mathcal{M}$ satisfying $\mathcal{P}$ (\ie
there exists a {\mc} $\mathcal{M}$ satisfying an {\imc} $\mathcal{I}$ instance of $\mathcal{P}$).
%
%
%
As seen in Section~\ref{sec:background}, {\pimc}s are equipped with two
semantics: once-and-for-all ($\satisfactionPimcOnce$) and
at-every-step ($\satisfactionPimc$). Recall that
$\satisfactionPimcOnce$ imposes that the underlying graph structure of
implementations needs to be isomorphic to the graph structure of the
corresponding specification. In contrast, $\satisfactionPimc$ allows
implementations to have a different graph structure. It therefore
seems that some {\pimc}s could be inconsistent w.r.t
$\satisfactionPimcOnce$ while being consistent w.r.t
$\satisfactionPimc$. On the other hand, checking the consistency w.r.t
$\satisfactionPimcOnce$ seems easier because of the fixed graph
structure.

In \cite{Delahaye15}, the author firstly proved that both semantics are equivalent
w.r.t. existential consistency, and proposed a {\csp} encoding for verifying this property
which is exponential in the size of the {\pimc}.
Based on this result of semantics equivalence
w.r.t. existential consistency from \cite{Delahaye15}
we propose a new {\csp} encoding, written
{\Mec}, for verifying the existential consistency property for
{\pimc}s.


\begin{figure*}[t]
\mbox{
\begin{minipage}[t]{.48\textwidth}
\begin{center}
{\tiny
\begin{tikzpicture}[scale=1.4]
	\node[vertex] (n0) at (0,0.5) {$\rho_0$};
	\node[vertex] (n1) at (1.2,1) {$\rho_1$};
	\node[vertex] (n2) at (1.2,0) {$\rho_2$};
	\node[vertex] (n3) at (2.4,1) {$\rho_3$};
	\node[vertex] (n4) at (2.4,0) {$\rho_4$};
	\node[vertex,draw=none,fill=none,minimum size=0em] (nP) at (3.5,0.66) {$\pi_p \in [0,1]$};
	\node[vertex,draw=none,fill=none,minimum size=0em] (nQ) at (3.5,0.33) {$\pi_q \in [0,1]$};

	\draw[edge] (n0) to node[above left] {$\transition{0}{1}$} (n1);
	\draw[edge] (n0) to node[below left] {$\transition{0}{2}$} (n2);

	\draw[edge] (n1) to[out=45,in=45+90,min distance=4mm] node[above] {$\transition{1}{1}$} (n1);
	\draw[edge] (n1) to node[above] {$\transition{1}{3}$} (n3);
	
	\draw[edge] (n2) to node[right] {$\transition{2}{1}$} (n1);
	\draw[edge] (n2) to[out=-45,in=-45-90,min distance=4mm] node[below] {$\transition{2}{2}$} (n2);
	\draw[edge] (n2) to node[below] {$\transition{2}{4}$} (n4);	
	
	\draw[edge] (n3) to[out=45,in=-45,min distance=4mm] node[right] {$\transition{3}{3}$} (n3);
	
	\draw[edge] (n4) to node[right] {$\transition{4}{3}$} (n3);
	\draw[edge] (n4) to[out=-45,in=45,min distance=4mm] node[right] {$\transition{4}{4}$} (n4);
\end{tikzpicture}}
\caption{Variables in the {\csp} produced by {\Mec} for the {\pimc} $\mathcal{P}$ from Fig. \ref{fig:example_pimc}}\label{fig:variables_consistency}
\end{center}
\end{minipage}}
\hspace{.005\textwidth}
\mbox{
\begin{minipage}[t]{.48\textwidth}
\begin{center}
{\tiny
\hspace{-0.4cm}
\begin{tikzpicture}[scale=1.4]
	\node[vertex] (n0) at (0,0.5) {$\top$};
	\node[vertex] (n1) at (1.2,1) {$\top$};
	\node[vertex] (n2) at (1.2,0) {$\top$};
	\node[vertex] (n3) at (2.4,1) {$\top$};
	\node[vertex] (n4) at (2.4,0) {$\bot$};
	\node[vertex,draw=none,fill=none,minimum size=0em] (nP) at (3.4,0.66) {$\pi_p = 0.5$};
	\node[vertex,draw=none,fill=none,minimum size=0em] (nQ) at (3.4,0.33) {$\pi_q = 0.5$};

	\draw[edge] (n0) to node[above left] {$0.7$} (n1);
	\draw[edge] (n0) to node[below left] {$0.3$} (n2);

	\draw[edge] (n1) to[out=45,in=45+90,min distance=4mm] node[above] {$0.5$} (n1);
	\draw[edge] (n1) to node[above] {$0.5$} (n3);
	
	\draw[edge] (n2) to node[right] {$0.5$} (n1);
	\draw[edge] (n2) to[out=-45,in=-45-90,min distance=4mm] node[below] {$0.5$} (n2);
	\draw[edge] (n2) to node[below] {$0$} (n4);	
	
	\draw[edge] (n3) to[out=45,in=-45,min distance=4mm] node[right] {$1$} (n3);
	
	\draw[edge] (n4) to node[right] {$0$} (n3);
	\draw[edge] (n4) to[out=-45,in=45,min distance=4mm] node[right] {$0$} (n4);
\end{tikzpicture}}
\caption{A solution to the {\csp} {\Mec}($\mathcal{P}$) for the {\pimc} $\mathcal{P}$ from Fig. \ref{fig:example_pimc}}\label{fig:solution_consistency}
\end{center}
\end{minipage}}
\vspace*{-0.4cm}
\end{figure*}

Let $\mathcal{P}$ $=$ $(S,$$s_0,$$P,$$V,$$Y)$ be a {\pimc}, we
write \Mec($\mathcal{P}$) for the {\csp} produced by {\Mec} according
to $\mathcal{P}$. Any solution of \Mec($\mathcal{P}$) will
  correspond to a {\mc} satisfying $\mathcal{P}$.  In
\Mec($\mathcal{P}$), we use one variable $\pi_p$ with domain $[0,1]$
per parameter $p$ in $Y$; one variable $\transition{s}{s^\prime}$ with
domain $[0, 1]$ per transition $(s, s^\prime)$ in $\{ \{s\} \times
\Succ(s) \mid s \in S\}$; and
one Boolean variable $\rho_s$ per state $s$ in $S$.
These Boolean variables will indicate for each state whether it appears in the {\mc} solution of the {\csp} (\ie in the {\mc} satisfying the {\pimc} $\mathcal{P}$).
For each state $s \in S$,  Constraints are as follows:

\begin{minipage}{\textwidth}
  \hspace{-.4cm}
  \begin{minipage}[t]{.54\textwidth}
    \begin{enumerate}
      \enumerateConstraints
      \item {$\rho_{s}$, 
         if $s = s_0$}\label{encoding_ec_init_state}%
      \setcounter{enumi}{2}
      \item {$\neg \rho_s \Leftrightarrow 
        \Sigma_{s^\prime \in \Pred(s) \setminus \{s\}} \transition{s^\prime}{s} = 0$,
         if $s \ne s_0$}\label{encoding_ec_cstr_reach_propag}%
      \setcounter{enumi}{4}
      \item { $\rho_s \Rightarrow 
        \transition{s}{s^\prime} \in P(s,s^\prime)$,
         for all $s^\prime \in \Succ(s)$}\label{encoding_ec_cstr_intervals}%
    \end{enumerate}
  \end{minipage}
  \hspace{.04\textwidth}
  \begin{minipage}[t]{.4\textwidth}
    \begin{enumerate}
      \enumerateConstraints
      \setcounter{enumi}{1}
      \item { $\rho_s \Leftrightarrow 
        \Sigma_{s^\prime \in \Succ(s)} \transition{s}{s^\prime} = 1$}\label{encoding_ec_sum_to_one}%
      \setcounter{enumi}{3}
      \item { $\neg \rho_s \Leftrightarrow 
        \Sigma_{s^\prime \in \Succ(s)} \transition{s}{s^\prime} = 0$}\label{encoding_ec_sum_to_zero}%
    \end{enumerate}
  \end{minipage}  
\end{minipage}

\smallskip

Recall that given a {\pimc} $\mathcal{P}$ the objective of
  the {\csp} $\Mec(\mathcal{P})$ is to construct a {\mc} $\mathcal{M}$
  satisfying $\mathcal{P}$.  Constraint~\ref{encoding_ec_init_state} states that the initial
  state $s_0$ appears in $\mathcal{M}$.  Constraint~\ref{encoding_ec_cstr_reach_propag} ensures
  that for each non-initial state $s$, variable $\rho_s$ is set to
  {\false} iff $s$ is not reachable from its predecessors.
  Constraint~\ref{encoding_ec_sum_to_one} ensures that if a
  state $s$ appears in $\mathcal{M}$, then its outgoing transitions
  form a probability distribution. 
  On the contrary, Constraint~\ref{encoding_ec_sum_to_zero} propagates 
  non-appearing states (\ie if a state $s$
  does not appear in $\mathcal{M}$ then all its outgoing transitions
  are set to zero).   Finally, Constraint~\ref{encoding_ec_cstr_intervals} states
  that, for all appearing states, the outgoing transition
  probabilities must be selected inside the specified intervals.

\begin{example}\label{ex:model_consistency} Consider the {\pimc} $\mathcal{P}$ given in Figure \ref{fig:example_pimc}.
Figure \ref{fig:variables_consistency} describes the variables in $\Mec(\mathcal{P})$:
one variable per transition 
(\eg $\transition{0}{1}$, $\transition{0}{2}$, $\transition{1}{1}$), 
one Boolean variable per state
(\eg $\rho_0$, $\rho_1$), 
and one variable per parameter ($\pi_p$ and $\pi_q$).
The following constraints
correspond to the Constraints~\ref{encoding_ec_sum_to_one}, \ref{encoding_ec_cstr_reach_propag}, \ref{encoding_ec_sum_to_zero}, and \ref{encoding_ec_cstr_intervals} generated by our encoding $\Mec$
for the state $2$ of $\mathcal{P}$:

\hspace*{-0.7cm}
\begin{minipage}[t]{.34\textwidth}
\begin{enumerate}
	\item[]	$\neg \rho_2 \Leftrightarrow \transition{0}{2}  = 0$
	\item[] $\neg \rho_2 \Leftrightarrow \transition{2}{1} + \transition{2}{2} + \transition{2}{4} = 0$
\end{enumerate}
\end{minipage}
\begin{minipage}[t]{.32\textwidth}
\begin{enumerate}
	\item[] $\rho_2 \Leftrightarrow \transition{2}{1} + \transition{2}{2} + \transition{2}{4} = 1$
    \item[] $\rho_2 \Rightarrow 0 \leq \transition{2}{1} \leq \pi_p$
\end{enumerate}
\end{minipage}
\begin{minipage}[t]{.32\textwidth}
\begin{enumerate}
	\item[] $\rho_2 \Rightarrow 0.2 \leq \transition{2}{2} \leq \pi_p$
	\item[] $\rho_2 \Rightarrow 0 \leq \transition{2}{4} \leq 0.5$
\end{enumerate}
\end{minipage}
\end{example}

Finally, Figure \ref{fig:solution_consistency} describes a solution for the {\csp} $\Mec(\mathcal{P})$.
Note that given a solution of a {\pimc} encoded by $\Mec$,
one can construct a {\mc} satisfying the given {\pimc}
by keeping all the states $s$ s.t. $\rho_s$ is equal to {\true} and
considering the transition function given by the probabilities in the
$\transition{s}{s^\prime}$ variables. We now show that
our encoding works as expected.

\begin{proposition}\label{prop:csp_existential_consistency}
	A {\pimc} $\mathcal{P}$ is existential consistent
    iff $\Mec(\mathcal{P})$ is satisfiable.
\end{proposition}

Our existential consistency encoding is linear in the size of the {\pimc} instead of exponential for the encoding from~\cite{DelahayeLP16} which enumerates the powerset of the states in the {\pimc}
resulting in deep nesting of conjunctions and disjunctions. 

\subsection{Qualitative Reachability}

Let $\mathcal{P} = (S,s_0,P,V,Y)$ be a \pimc\ and 
$\alpha \subseteq A$ be a state label.
We say that $\alpha$ is {\em existential reachable} in $\mathcal{P}$ 
iff there exists an implementation $\mathcal{M}$ of $\mathcal{P}$
where $\alpha$ is reachable
(\ie $\Proba^{\mathcal{M}}(\ltlExists \alpha)>0$).
In a dual way, 
we say that $\alpha$ is {\em universal reachable} in $\mathcal{P}$ 
iff $\alpha$ is reachable in any implementation $\mathcal{M}$ of $\mathcal{P}$.
As for existential consistency, we use a result from~\cite{Delahaye15}
that states that both {\pimc} semantics are equivalent w.r.t.
existential (and universal) reachability. We therefore
propose a new CSP encoding, written $\Mer$, that extends $\Mec$, for verifying these properties. Formally, 
{\csp} $\Mer(\mathcal{P}) = (X \cup X^\prime,D \cup D^\prime,C \cup C^\prime)$ is such that
$(X,D,C) = \Mec(\mathcal{P})$, 
$X^\prime$~contains one integer variable $\omega_s$  with domain $[0, |S|]$ per state $s$ in $S$,
$D^\prime$~contains the domains of these variables, and 
$C^\prime$ is composed of the following constraints for each state $s \in S$:
\smallskip
\begin{minipage}{\textwidth}
  \hspace*{.2cm}
  \begin{minipage}[t]{.3\textwidth}
    \begin{enumerate}
      \enumerateConstraints
      \setcounter{enumi}{5}
      \item {$\omega_{s} = 1$, if $s = s_0$}\label{encoding_er_init_state}%
    \end{enumerate}
  \end{minipage}
  \hspace{.02\textwidth}
  \begin{minipage}[t]{.3\textwidth}
    \begin{enumerate}
      \enumerateConstraints
      \setcounter{enumi}{6}
      \item { $\omega_s \neq 1$, if $s \neq s_0$}\label{encoding_er_non_init_state}%
    \end{enumerate}
  \end{minipage}
  \hspace{.02\textwidth}
  \begin{minipage}[t]{.28\textwidth}
    \begin{enumerate}
      \enumerateConstraints
      \setcounter{enumi}{7}
      \item { $\rho_s \Leftrightarrow (\omega_s \neq 0)$}\label{encoding_er_bool_var}%
    \end{enumerate}
  \end{minipage}  
  \hspace*{0.25cm}
  \begin{minipage}[t]{.93\textwidth}
  	\vspace*{-0.25cm}
    \begin{enumerate}
      \enumerateConstraints
      \setcounter{enumi}{8}
      \item { $\omega_s > 1 \Rightarrow \bigvee_{s^\prime \in \Pred(s) \setminus \{ s \} }(\omega_s = \omega_{s^\prime} + 1) \wedge (\transition{s}{s^\prime} > 0)$,
		 if $s \neq s_0$}\label{encoding_er_propag_reach}%
      \item { $\omega_s = 0 \Leftrightarrow \bigwedge_{s^\prime \in \Pred(s) \setminus \{ s \} }(\omega_{s^\prime} = 0) \vee (\transition{s}{s^\prime} = 0)$,
		 if $s \neq s_0$}\label{encoding_er_propag_non_reach}%
    \end{enumerate}
  \end{minipage}  
\end{minipage}

\smallskip


Recall first that {\csp} $\Mec(P)$ constructs a Markov chain $\mathcal{M}$ satisfying $\mathcal{P}$.
Informally, for each state $s$ in $\mathcal{M}$ the Constraints~\ref{encoding_er_init_state}, \ref{encoding_er_non_init_state}, \ref{encoding_er_propag_reach} and \ref{encoding_er_propag_non_reach} in $\Mer$ ensure that
$\omega_s = k$ iff there exists in $\mathcal{M}$
a path from the initial state to $s$ of length $k-1$ with non zero probability;
and state $s$ is not reachable in $\mathcal{M}$ from the initial state $s_0$ iff $\omega_s$ equals to $0$. Finally, Constraint~\ref{encoding_er_bool_var} enforces the Boolean reachability indicator variable $\rho_s$
to bet set to {\true} iff there exists a path with non zero probability in $\mathcal{M}$ 
from the initial state $s_0$ to $s$ (\ie $\omega_s \neq
0$).

Let $S_\alpha$ be the set of states from $\mathcal{P}$ labeled with $\alpha$.
$\Mer(\mathcal{P})$ therefore produces a Markov chain satisfying
$\mathcal{P}$ where reachable states $s$ are such that $\rho_s =
\true$.
As a consequence, $\alpha$ is existential reachable in
$\mathcal{P}$ iff $\Mer(\mathcal{P})$ admits a solution such that
$\bigvee_{s \in S_\alpha} \rho_s$; and $\alpha$ is universal reachable in
$\mathcal{P}$ iff $\Mer(\mathcal{P})$ admits no solution such that
$\bigwedge_{s \in S_\alpha} \neg\rho_s$. This is formalised in the following
proposition.

\begin{proposition}\label{prop:model_existential_reachability}
	Let $\mathcal{P} = (S, s_0 , P, V, Y)$ be a \pimc, 
    $\alpha \subseteq A$ be a state label, 
    $S_\alpha = \{s \ | \ V(s) = \alpha\}$,
    and $(X,D,C)$ be the {\csp} $\Mer(\mathcal{P})$.
    \vspace*{-0.15cm}
    \begin{itemize}
    	\item 
			\csp\ $(X,D,C \cup \bigvee_{s \in S_\alpha} \rho_s)$
    		is satisfiable iff 
    		$\alpha$ is existential reachable in $\mathcal{P}$
   		\item
			\csp\ $(X,D,C \cup \bigwedge_{s \in S_\alpha} \neg\rho_s)$
    		is unsatisfiable iff 
    		$\alpha$ is universal reachable in $\mathcal{P}$
	\end{itemize}
\end{proposition}

As for the existential consistency problem, 
we have an exponential gain in terms of size of the encoding
compared to~\cite{DelahayeLP16}:
the number of constraints and variables in {\Mer} is linear
in terms of the size of the encoded {\pimc}.

\custompar{Remark.}
In $\Mer$ Constraints~\ref{encoding_ec_cstr_reach_propag} inherited from $\Mec$
are entailed by Constraints~\ref{encoding_er_bool_var} and \ref{encoding_er_propag_non_reach} added to $\Mer$. 
Thus, in a practical approach one may ignore Constraints~\ref{encoding_ec_cstr_reach_propag} from $\Mec$ 
if they do not improve the solver performances.

\section{Quantitative Properties}\hspace*{0cm}
\label{sec:quantitative}

We now move to the verification of quantitative reachability
properties in {\pimc}s. Quantitative reachability has already been
investigated in the context of {\pmc}s and {\imc}s with the
once-and-for-all semantics. Due to the complexity of allowing
implementation structures to differ from the structure of the
specifications, quantitative reachability in {\imc}s with the
at-every-step semantics has, to the best of our knowledge, never been
studied. In this section, we propose our main theoretical contribution: a theorem
showing that both {\imc} semantics are equivalent with respect to
quantitative reachability, which allows the extension of all results
from~\cite{tulip,benedikt2013ltl} to the at-every-step semantics. Based on this result, we
also extend the CSP encodings introduced in
Section~\ref{sec:qualitative-reachability} in order to solve quantitative
reachability properties on {\pimc}s regardless of their semantics.


\subsection{Equivalence of $\satisfactionImcOnce$ and
  $\satisfactionImc$ w.r.t quantitative reachability}\label{sec:equiv_imc_semantics}

Given an {\imc} $\mathcal{I} = (S,s_0,P,V)$ and a state label
$\alpha \subseteq A$, a quantitative reachability property on
$\mathcal{I}$ is a property of the type
$\mathbb{P}^{\mathcal{I}}(\ltlExists \alpha) {\sim} p$, where
$0<p<1$ and ${\sim} \in \{\le, <, >, \ge\}$. Such a property is
verified iff there exists an {\mc} $\mathcal{M}$ satisfying $\mathcal{I}$ (with the chosen semantics) such that
$\mathbb{P}^{\mathcal{M}}(\ltlExists \alpha) {\sim} p$.

As explained above, all existing techniques and tools for verifying
quantitative reachability properties on {\imc}s only focus on the
once-and-for-all semantics. Indeed, in this setting, quantitative
reachability properties are easier to compute because the underlying
graph structure of all implementations is known. However, to the best
of our knowledge, there are no works addressing the same problem with
the at-every-step semantics or showing that addressing the problem in
the once-and-for-all setting is sufficiently general. The following
theorem fills this theoretical gap by proving that both semantics are
equivalent w.r.t quantitative reachability. In other words, for all
{\mc} $\mathcal{M}$ such that $\mathcal{M} \satisfactionImc
\mathcal{I}$ and all state label $\alpha$, there exist {\mc}s
$\mathcal{M}_\le$ and $\mathcal{M}_{\ge}$ such that $\mathcal{M}_{\le}
\satisfactionImcOnce \mathcal{I}$, $\mathcal{M}_{\ge}
\satisfactionImcOnce \mathcal{I}$ and
$\mathbb{P}^{\mathcal{M}_{\le}}(\ltlExists \alpha) \le
\mathbb{P}^{\mathcal{M}}(\ltlExists \alpha) \le
\mathbb{P}^{\mathcal{M_{\ge}}}(\ltlExists \alpha)$. This is
formalized in the following theorem.

\begin{theorem}\label{thm:reachability-semantics-equivalence-imcs}
	Let $\mathcal{I} = (S,s_0,P,V)$ be an {\imc}, $\alpha
        \subseteq A$ be a state label, ${\sim} \in \{\le,<,>,\ge\}$
        and $0<p<1$. $\mathcal{I}$ satisfies
        $\mathbb{P}^{\mathcal{I}}(\ltlExists \alpha) {\sim} p$ with
        the once-and-for-all semantics iff $\mathcal{I}$ satisfies
        $\mathbb{P}^{\mathcal{I}}(\ltlExists \alpha) {\sim} p$ with
        the at-every-step semantics.
\end{theorem}

The proof is constructive (see 
Appendix~\ref{ap:equiv_imc_semantics}): we use the structure of the relation $\mathcal{R}$ from the
definition of $\satisfactionImc$ in order to build the {\mc}s
$\mathcal{M}_{\le}$ and $\mathcal{M}_{\ge}$.

\subsection{Constraint Encodings}

Note that the result from
Theorem~\ref{thm:reachability-semantics-equivalence-imcs} naturally
extends to {\pimc}s. We therefore exploit this result to construct a
{\csp} encoding for verifying quantitative reachability properties in
{\pimc}s.
As in Section~\ref{sec:qualitative-reachability}, we extend the
CSP $\Mec$, that produces a correct $\mc$ implementation for the given
{\pimc}, by imposing that this $\mc$ implementation satisfies the
given quantitative reachability property. In order to compute the
probability of reaching state label $\alpha$ at the {\mc} level, we
use standard techniques from~\cite{Baier2008PMC} that require the
partitioning of the state space into three sets $S_{\top}$,
$S_{\bot}$, and $S_?$ that correspond to states reaching
$\alpha$ with probability $1$, states from which $\alpha$ cannot be
reached, and the remaining states, respectively. Once this partition is chosen, the
reachability probabilities of all states in $S_?$ are computed as the
unique solution of a linear equation system (see~\cite{Baier2008PMC},
Theorem 10.19, p.766). We now explain how we identify states from
$S_\bot, S_\top$ and $S_?$ and how we encode the
linear equation system, which leads to the resolution of quantitative
reachability.

Let $\mathcal{P} = (S,s_0,P,V,Y)$ be a \pimc\ and $\alpha \subseteq
A$ be a state label. We start by setting $S_\top = \{s \ |\ V(s) =
\alpha\}$. We then extend $\Mer(\mathcal{P})$ in order to identify the
set $S_\bot$. Let $\MerPrime(\mathcal{P}, \alpha) = (X \cup X^\prime,D
\cup D^\prime,C \cup C^\prime)$ be such that $(X,D,C) =
\Mer(\mathcal{P})$, $X^\prime$~contains one Boolean variable
$\lambda_s$ and one integer variable $\alpha_s$ with domain $[0, |S|]$
per state $s$ in $S$, $D^\prime$~contains the domains of these
variables, and $C^\prime$ is composed of the following constraints for
each state $s \in S$:

\begin{minipage}{\textwidth}
  \hspace{-0.4cm}
  \begin{minipage}[t]{.3\textwidth}
    \begin{enumerate}
      \enumerateConstraints
      \setcounter{enumi}{10}
      \item {$\alpha_s = 1$,
	 if $\alpha = V(s)$}\label{encoding_erprime_target_state}%
    \end{enumerate}
  \end{minipage}
  \hspace*{0.02\textwidth}
  \begin{minipage}[t]{.3\textwidth}
    \begin{enumerate}
      \enumerateConstraints
      \setcounter{enumi}{11}
      \item { $\alpha_s \neq 1$,
	 if $\alpha \ne V(s)$}\label{encoding_erprime_non_target_state}%
    \end{enumerate}
  \end{minipage}
  \hspace*{0.02\textwidth}
  \begin{minipage}[t]{.34\textwidth}
    \begin{enumerate}
      \enumerateConstraints
      \setcounter{enumi}{12}
      \item { $\lambda_s \Leftrightarrow (\rho_s \wedge (\alpha_s \neq 0))$}\label{encoding_erprime_bool_var}%
    \end{enumerate}
  \end{minipage}  
  \hspace*{-0.4cm}
  \begin{minipage}[t]{0.99\textwidth}
  	\vspace*{-0.2cm}
    \begin{enumerate}
      \enumerateConstraints
      \setcounter{enumi}{13}
      \item { $\alpha_s > 1 \Rightarrow \bigvee_{s^\prime \in \Succ(s) \setminus \{ s \} }(\alpha_s = \alpha_{s^\prime} + 1) \wedge (\transition{s}{s^\prime} > 0)$,
	 if $\alpha \ne V(s)$}\label{encoding_erprime_propag_target}%
      \item { $\alpha_s = 0 \Leftrightarrow \bigwedge_{s^\prime \in \Succ(s) \setminus \{ s \} }(\alpha_{s^\prime} = 0) \vee (\transition{s}{s^\prime} = 0)$,
	 if $\alpha \ne V(s)$}\label{encoding_erprime_propag_non_target}%
    \end{enumerate}
  \end{minipage}  
\end{minipage}

\smallskip


Note that  variables $\alpha_s$ play a symmetric role to 
variables $\omega_s$ from $\Mer$: instead of indicating the existence
of a path from $s_0$ to $s$, they characterize the existence of a path
from $s$ to a state labeled with $\alpha$. In addition, due to
Constraint~\ref{encoding_erprime_bool_var},  variables $\lambda_s$ are set to {\true} iff
there exists a path with non zero probability from the initial state
$s_0$ to a state labeled with $\alpha$ passing by $s$. Thus, $\alpha$ cannot be reached from states s.t.
$\lambda_s = \false$.
Therefore, $S_\bot = \{s \ |\ \lambda_s = \false\}$.

Finally, we encode the equation system from~\cite{Baier2008PMC} in a
last {\csp} encoding that extends $\MerPrime$. Let
$\MerExt(\mathcal{P}, \alpha) = (X \cup X^\prime,D \cup D^\prime,C
\cup C^\prime)$ be such that $(X,D,C) = \MerPrime(\mathcal{P},
\alpha)$, $X^\prime$~contains one variable $\pi_s$ per state $s$ in
$S$ with domain $[0, 1]$, $D^\prime$~contains the domains of these
variables, and $C^\prime$ is composed of the following constraints for
each state $s \in S$:

\begin{minipage}{0.93\textwidth}
  \begin{minipage}[t]{.45\textwidth}
    \begin{enumerate}
      \enumerateConstraints
      \setcounter{enumi}{15}
      \item {$\neg\lambda_s \Rightarrow \pi_s = 0$}%
    \end{enumerate}
  \end{minipage}
  \begin{minipage}[t]{.45\textwidth}
    \begin{enumerate}
      \enumerateConstraints
      \setcounter{enumi}{16}
      \item { $\lambda_s \Rightarrow \pi_s = 1$,
	if $\alpha = V(s)$}%
    \end{enumerate}
  \end{minipage}
  \begin{minipage}[t]{0.67\textwidth}
    \begin{enumerate}
      \enumerateConstraints
      \setcounter{enumi}{17}
      \item { $ 
		\lambda_s \Rightarrow \pi_s = \Sigma_{s^{\prime} \in \Succ(s)} \pi_{s^\prime} \transition{s^\prime}{s}$,
		\hfill if $\alpha \ne V(s)$}%
    \end{enumerate}
  \end{minipage}  
\end{minipage}

\smallskip


As a consequence, variables $\pi_s$ encode the probability with which
state $s$ eventually reaches $\alpha$ when $s$ is reachable from the
initial state and $0$ otherwise.

Let $p \in [0, 1] \subseteq \Rset$ be a probability bound.  Adding the
constraint $\pi_{s_0} \leq p$ (resp. $\pi_{s_0} \geq p$) to the
previous {\MerExt} encoding allows to determine if there exists a
{\mc} $\mathcal{M} \satisfactionPimc \mathcal{P}$ such that
$\mathbb{P}^{\mathcal{M}} (\ltlExists \alpha) \le p$ (resp $\ge p$).
Formally, let ${\sim} \in \{\leq, <, \geq, >\}$ be a comparison operator,
we write $\not\sim$ for its negation (\eg $\not\leq$ is $>$).
This leads to the following theorem.

\begin{theorem}\label{thm:pimc_reachability_in_cp}
	Let $\mathcal{P} = (S, s_0 , P, V, Y)$ be a \pimc,
	$\alpha \subseteq A$ be a label,
	$p \in [0, 1]$,
    ${\sim} \in \{\leq,<, \geq,>\}$ be a comparison operator,
    and $(X,D,C)$ be $\MerExt(\mathcal{P}, \alpha)$:
    \vspace*{-0.05cm}
    \begin{itemize}
    	\item 
			\csp\ $(X,D,C \cup (\pi_{s_0} \sim p))$
    		is satisfiable iff 
    		$\exists \mathcal{M} \satisfactionPimc \mathcal{P}$ s.t. $\Proba^\mathcal{M}(\ltlExists \alpha) \sim p$
   		\item
			\csp\ $(X,D,C \cup (\pi_{s_0} \not\sim p))$
    		is unsatisfiable iff 
    		$\forall \mathcal{M} \satisfactionPimc \mathcal{P}$: $\Proba^\mathcal{M}(\ltlExists \alpha) \sim p$
	\end{itemize}
\end{theorem}

\section{Prototype Implementation}

Our results have been implemented in a prototype tool\footnote{All
  resources, benchmarks, and source code  are
  available online as a Python library at \url{https://github.com/anicet-bart/pimc_pylib}}
which generates  the above CSP encodings, and
CSP encodings from~\cite{DelahayeLP16} as well. Given a {\pimc} in a text
format inspired from \cite{tulip},
our tool produces the desired {\csp} as a SMT instance with the QF\_NRA logic (Quantifier Free Non linear Real-number
Arithmetic). This instance can then be fed to any
solver accepting the SMT-LIB format with QF\_NRA logic \cite{BarFT-SMTLIB}. For our
benchmarks, we chose Z3 \cite{Z3}  (latest version: 4.5.0).

QF\_NRA does not deal with integer variables. In practice,
logics mixing integers and reals are harder than those over reals only. Thus we obtained better results by encoding integer
variables into real ones. In our implementations each integer variable $x$ is declared as a real variable whose real domain bounds are its original integer domain bounds; we also add the constraint $x < 1 \Rightarrow x = 0$. Since we only perform incrementation of $x$ this preserves the  same set of solutions. 

In order to evaluate our prototype, we extend the 
\bench{nand} model from \cite{NPKS05}\footnote{Available online at
  \texttt{http://www.prismmodelchecker.com}}. The original {\mc}
\bench{nand} model has already been extended as a {\pmc}
in~\cite{Prophesy}, where the authors consider a single parameter $p$
for the probability that each of the $N$ $nand$ gates fails during the
multiplexing. We extend this model to {\pimc} by considering intervals
for the probability that the initial inputs are stimulated and we
have one parameter per $nand$ gate to represent the
probability that it fails.  {\pimc}s in text format are
automatically generated from the PRISM model.

Table~\ref{tab:xp} summarizes the size of the considered instances of
the model (in terms of states, transitions, and parameters) and of the
corresponding CSP problems (in terms of number of variables and
constraints). In addition, we also present the resolution time of the
given CSPs using the Z3 solver. Our experiments were performed on a $2.4$ GHz Intel Core
i5 processor with time out set to $10$ minutes and memory out set to
$2$Gb.

\begin{table}[t]
  {  \scriptsize
    \begin{center}
  \scalebox{.9}{\begin{tabular}{|ll||ccc|ccc|ccc|ccc|c|}
\hline
   &  & \multicolumn{3}{c|}{\pimc} & \multicolumn{3}{c|}{\Mec} & \multicolumn{3}{c|}{\Mer}
   & \multicolumn{3}{c|}{\MerExt} \\
  \multicolumn{2}{|l||}{Benchmark} & 
  \#states & \#trans. & \#par. &
  \#var. & \#cstr. & time &
  \#var. & \#cstr. & time &
  \#var. & \#cstr. & time \\
  \hline \hline

\bench{nand}      & \bench{K=1; N=2} & 104    & 147    & 4  & 255    & 1,526   & 0.17s & 170   & 1,497  & 0.19s  & 296    & 2,457   & 69.57s \\
\bench{nand}      & \bench{K=1; N=3} & 252    & 364    & 5  & 621    & 3,727   & 0.24s & 406   & 3,557  & 0.30s  & 703    & 5,828   & 31.69s \\
\bench{nand}      & \bench{K=1; N=5} & 930    & 1,371  & 7  & 2,308  &
13,859  & 0.57s & 1,378 & 12,305 & 0.51s   & 2,404 & 20,165  & T.O. \\
\bench{nand}      & \bench{K=1; N=10} & 7,392 & 11,207 & 12 & 18,611 & 111,366 & 9.46s & 9,978 & 89,705 & 13.44s & 17,454 & 147,015 & T.O. \\\hline
  \end{tabular}}
  \end{center}}
\caption{Benchmarks}\label{tab:xp}\vspace{-1cm}
\end{table}

\section{Conclusion and future work}
In this paper, we have compared several Markov Chain abstractions in
terms of succinctness and we have shown that Parametric Interval Markov Chain
is a strictly more succinct abstraction formalism than other existing
formalisms such as Parametric Markov Chains and Interval Markov
Chains. In addition, we have proposed constraint encodings for
checking several properties over {\pimc}. In the context of
qualitative properties such as existencial consistency or consistent
reachability, the size of our encodings is significantly smaller than
other existing solutions. In the quantitative setting, we have
compared the two usual semantics for {\imc}s and {\pimc}s and showed
that both semantics are equivalent with respect to quantitative
reachability properties. As a side effect, this result ensures that
all existing tools and algorithms solving reachability problems in
{\imc}s under the once-and-for-all semantics can safely be extended to
the at-every-step semantics with no changes. Based on this result, we have then proposed
{\csp} encodings addressing quantitative reachability in the context of
{\pimc}s regardless of the chosen semantics. Finally, we have
developed a prototype tool that automatically generates our {\csp}
encodings and that can be plugged to any constraint solver accepting
the SMT-LIB format as input.

We plan to develop our tool for  {\pimc}
verification in order to manage other, more complex, properties (\eg
supporting the LTL-language in the spirit of what Tulip \cite{tulip}
does).  We also plan on investigating a practical way of computing and
representing the set of {\em all solutions} to the parameter synthesis
problem.

\bibliographystyle{splncs03}
\bibliography{biblio}

\newpage

\appendix

\section{Complements to Section~\ref{sec:abstraction-models}}

\subsection{$\satisfactionImc$ is More General
than $\satisfactionImcOnce$}\label{ap:compare_imcs_satisfaction_relations}

The {\em at-every-step} satisfaction relation is more general than the {\em once-and-for-all} satisfaction relation. Let $\mathcal{I} = (S, s_0, P, V)$ be an {\imc} and
$\mathcal{M} = (T, t_0, p, V^\prime)$ be a {\mc}.
We show that 
1. $\mathcal{M} \satisfactionImcOnce \mathcal{I} \Rightarrow \mathcal{M} \satisfactionImc \mathcal{I}$;
2. in general $\mathcal{M} \satisfactionImc \mathcal{I} \not\Rightarrow \mathcal{M}~\satisfactionImcOnce~\mathcal{I}$.
The following examples also illustrates that even if a Markov chain satisfies an {\imc}
with the same graph representation w.r.t. the {\satisfactionImc} relation 
it may not verify the {\satisfactionImcOnce} relation.

\begin{figure}[t]
\centering
\mbox{
\begin{minipage}[t]{.28\textwidth}
\centering
{\tiny
\hspace*{0.4cm}
\begin{tikzpicture}[scale=1.4]
	\node[vertex, label={[label distance=0cm]0 :$\emptyset$}](n0) at (1.2,1) {$s_0$};
	\node[vertex, label={[label distance=0cm]180 :$\alpha$}] (n1) at (0.6,0)  {$s_1$};
	\node[vertex, label={[label distance=0cm]0 :$\alpha$}] (n2) at (1.8,0)  {$s_2$};

	\draw[edge] (n0) to node[left]  {$[0,0.4]$} (n1);
    \draw[edge] (n0) to node[right] {$[0.6,1]$} (n2);
	\draw[edge] (n1) to[out=-45,in=-45-90,min distance=4mm] node[below] {$[0,1]$} (n1);
	\draw[edge, bend right] (n1) to node[below] {$[0,1]$} (n2);
	\draw[edge] (n2) to[out=-45-90,in=-45,min distance=4mm] node[below] {$[0,1]$} (n2);
	\draw[edge, bend right] (n2) to node[above] {$[0,1]$} (n1);
\end{tikzpicture}
}
{{\newline}{\imc} $\mathcal{I}$}
\end{minipage}}
\mbox{
\begin{minipage}[t]{.26\textwidth}
\centering
{\tiny
\hspace*{0.1cm}
\begin{tikzpicture}[scale=1.4]
	\node[vertex, label={[label distance=0cm]0 :$\emptyset$}](n0) at (1.2,1) {$t_0$};
	\node[vertex, label={[label distance=0cm]180 :$\alpha$}] (n1) at (0.6,0)  {$t_1$};
	\node[vertex, label={[label distance=0cm]0 :$\alpha$}] (n2) at (1.8,0)  {$t_2$};

	\draw[edge] (n0) to node[left]  {$0.4$} (n1);
    \draw[edge] (n0) to node[right] {$0.6$} (n2);
	\draw[edge] (n1) to[out=-45,in=-45-90,min distance=4mm] node[below] {$0.5$} (n1);
	\draw[edge] (n2) to[out=-45-90,in=-45,min distance=4mm] node[below] {$0.5$} (n2);
    \draw[edge, bend right] (n1) to node[below] {$0.5$} (n2);
    \draw[edge, bend right] (n2) to node[above] {$0.5$} (n1);
\end{tikzpicture}
}
{{\newline}{\mc} $\mathcal{M}_1$}
\end{minipage}}
\mbox{
\begin{minipage}[t]{.27\textwidth}
\centering
{\tiny
\hspace*{0.2cm}
\begin{tikzpicture}[scale=1.4]
	\node[vertex, label={[label distance=0cm]0 :$\emptyset$}](n0) at (1.2,1) {$t_0$};
	\node[vertex, label={[label distance=0cm]180 :$\alpha$}] (n1) at (0.6,0)  {$t_1$};
	\node[vertex, label={[label distance=0cm]0 :$\alpha$}] (n2) at (1.8,0)  {$t_2$};

	\draw[edge] (n0) to node[left]  {$0.5$} (n1);
    \draw[edge] (n0) to node[right] {$0.5$} (n2);
	\draw[edge] (n1) to[out=-45,in=-45-90,min distance=4mm] node[below] {$0.5$} (n1);
	\draw[edge] (n2) to[out=-45-90,in=-45,min distance=4mm] node[below] {$0.5$} (n2);
    \draw[edge, bend right] (n1) to node[below] {$0.5$} (n2);
    \draw[edge, bend right] (n2) to node[above] {$0.5$} (n1);
\end{tikzpicture}
}
{{\newline}{\mc} $\mathcal{M}_2$}
\end{minipage}}
\captionsetup{justification=centering}
\caption{
	{\imc} $\mathcal{I}$, {\mc}s $\mathcal{M}_1$ and $\mathcal{M}_2$ s.t.\break
    $\mathcal{M}_1 \satisfactionImc \mathcal{I}$ and 
    $\mathcal{M}_1 \satisfactionImcOnce \mathcal{I}$;
    $\mathcal{M}_2 \satisfactionImc \mathcal{I}$ and 
    $\mathcal{M}_2 \not\satisfactionImcOnce \mathcal{I}$ 
    \break
    the graph representation of $\mathcal{I}$, $\mathcal{M}_1$, and $\mathcal{M}_2$ are isomorphic;
	\label{fig:ex_imcs_more_general_semantic}
}%
\end{figure}
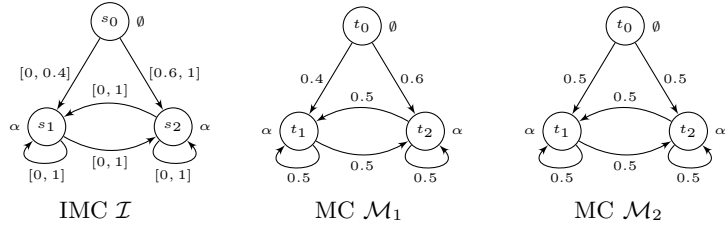

\begin{enumerate}
	\item 
If $\mathcal{M} \satisfactionImcOnce \mathcal{I}$ then by definition of $\satisfactionImcOnce$ we have that $T = S$, $t_0 = s_0$, and $p(s)(s^\prime) \in P(s,s^\prime)$. The relation $\mathcal{R} = \{ (s,s) \mid s \in S\}$ is a satisfaction relation between $\mathcal{M}$ and $\mathcal{I}$ (consider for each state in $S$ the correspondence function $\delta : S \to (S \to [0,1])$ s.t. $\delta(s)(s^\prime) = 1$ if $s = s^\prime$ and $\delta(s)(s^\prime) = 0$ otherwise).
Thus $\mathcal{M} \satisfactionImc \mathcal{I}$.

	\item Consider {\imc} $\mathcal{I}$, {\mc}s $\mathcal{M}_1$, $\mathcal{M}_2$, and $\mathcal{M}_3$ from Figure~\ref{fig:ex_imcs_more_general_semantic}.
    \begin{enumerate}
    	\item By definition of $\satisfactionImcOnce$ we have that $\mathcal{M}_1 \satisfactionImcOnce \mathcal{I}$. Thus by the previous remark we have that 
        $\mathcal{M}_1 \satisfactionImc \mathcal{I}$ ($t_0 = s_0$, $t_1 = s_1$ and $t_2 = s_2$).
        
        \item By definition of $\satisfactionImcOnce$ we have that $\mathcal{M}_2 \not\satisfactionImcOnce \mathcal{I}$. However the relation $\mathcal{R}$ containing
        $(t_0,s_0)$, $(t_1,s_1)$, $(t_1,s_2)$, $(t_2,s_1)$ and $(t_2,s_2)$ is a satisfaction relation between $\mathcal{I}$ and $\mathcal{M}_2$. Consider the functions: $\delta$ from $T$ to $(S \to [0,1])$
        s.t. $\delta(t_1)(s_1) = 4/5$, $\delta(t_1)(s_2) = 1/5$, 
        $\delta(t_2)(s_2) = 1$, and $\delta(t)(s) = 0$ otherwise;
        $\delta^\prime$ with the same signature than $\delta$
        s.t. $\delta^\prime(t_1)(s_1) = 1$, $\delta^\prime(t_2)(s_2) = 1$, 
        and $\delta^\prime(t)(s) = 0$ otherwise.
        We have that $\delta$ is a correspondence function for the pair $(t_0,s_0)$
        and $\delta^\prime$ is a correspondence function for all the remaining pairs in $\mathcal{R}$.
	Thus there exists a {\mc}s $\mathcal{M}$ s.t. $\mathcal{M} \satisfactionImc \mathcal{I}$
    and $\mathcal{M} \not\satisfactionImcOnce \mathcal{I}$.
    \end{enumerate}    
\end{enumerate}

    \subsection{Model Comparison}\label{ap:model_comparison}

    According to the given succinctness definition, 
 Lemma~\ref{lem:compare_imc_and_pmc} states that
 {\imcSet} and {\pmcSet} are not comparable w.r.t. both satisfaction relation $\satisfactionImcOnce$ and $\satisfactionImc$
 and that both satisfaction relations for {\imc}s are not comparable.

\begin{lemma}\label{lem:compare_imc_and_pmc}
	$\pmcSet$ and $\imcSet$ abstraction models are not comparable:
    \begin{enumerate*}
    	\item $\pmcSet \not\leq$ $(\imcSet, \satisfactionImc)$;
        \item $\pmcSet \not\leq (\imcSet, \satisfactionImcOnce)$;
        \item $(\imcSet, \satisfactionImc) \not\leq$ $\pmcSet$;
    	\item $(\imcSet, \satisfactionImc) \not\leq (\imcSet, \satisfactionImcOnce)$;
		\item $(\imcSet, \satisfactionImcOnce) \not\leq \pmcSet$;
    	\item $(\imcSet, \satisfactionImcOnce) \not\leq (\imcSet, \satisfactionImc)$;
	\end{enumerate*}
\end{lemma}

	We give a sketch of proof for each statement.
    Let $(\aminstance_1, \models_1)$ and $(\aminstance_2, \models_2)$ be two Markov chain abstraction models.
    Recall that according to the succinctness definition (\cf Definition~\ref{def:succinctness})
	$\aminstance_1 \not\leq \aminstance_2$
    if there exists
    $\mathcal{L}_2 \in \aminstance_2$ s.t.
    $\mathcal{L}_1 \not\equiv \mathcal{L}_2$
    for all $\mathcal{L}_1 \in \aminstance_1$.
    
\begin{figure}[t]
\mbox{
\begin{minipage}[t]{.12\textwidth}
\begin{center}
{\tiny
\begin{tikzpicture}[scale=1.4]

	\node[vertex, label={[label distance=-0.03cm]-45 :$\alpha$}](n1) at (1.2,1) {$1$};
	\node[vertex, label={[label distance=0cm]+180 :$\beta$}] (n0) at (1.2,0)  {$0$};

	\draw[edge] (n1) to[out=45,in=45+90,min distance=4mm] node[above] {$[0,1]$} (n1);
	\draw[edge] (n1) to node[left] {$[0,1]$} (n0);
	\draw[edge] (n0) to[out=-45,in=-45-90,min distance=4mm] node[below] {$1$} (n0);
\end{tikzpicture}
}
{{\imc} $\mathcal{I}$}\label{fig:imc_example_succinctness}
\end{center}
\end{minipage}}
\hspace{.005\textwidth}
\mbox{
\begin{minipage}[t]{.12\textwidth}
\begin{center}
{\tiny
\hspace{-0.4cm}
\begin{tikzpicture}[scale=1.3]

	\node[vertex, label={[label distance=-0.03cm]-45 :$\alpha$}](n1) at (1.2,1) {$s_1$};
	\node[vertex, label={[label distance=0cm]+180 :$\beta$}] (n0) at (1.2,0)  {$s_0$};

	\draw[edge] (n1) to[out=45,in=45+90,min distance=4mm] node[above] {$p$} (n1);
	\draw[edge] (n1) to node[left] {$1{-}p$} (n0);
	\draw[edge] (n0) to[out=-45,in=-45-90,min distance=4mm] node[below] {$1$} (n0);
\end{tikzpicture}
}
{{\pmc} $\mathcal{P}_1$}
\end{center}
\end{minipage}}
\hspace{.005\textwidth}
\mbox{
\begin{minipage}[t]{.21\textwidth}
\begin{center}
{\tiny
\hspace{-0.4cm}
\begin{tikzpicture}[scale=1.3]

	\node[vertex, label={[label distance=-0.03cm]-45 :$\alpha$}](n1) at (1.2,1) {$s_1$};
    \node[vertex, label={[label distance=-0.03cm]-45 :$\alpha$}](n2) at (2.4,1) {$s_2$};
	\node[vertex, label={[label distance=0cm]+180 :$\beta$}] (n0) at (1.8,0)  {$s_0$};

	\draw[edge] (n1) to[out=45,in=45+90,min distance=4mm] node[above] {$p_1$} (n1);
	\draw[edge] (n1) to node[below left] {$q_1$} (n0);
    \draw[edge] (n1) to node[above] {$1{-}p_1{-}q_1$} (n2);
	\draw[edge] (n2) to[out=45,in=45+90,min distance=4mm] node[above] {$p_2$} (n2);
	\draw[edge] (n2) to node[below right] {$1{-}p_2$} (n0);
    
	\draw[edge] (n0) to[out=-45,in=-45-90,min distance=4mm] node[below] {$1$} (n0);
\end{tikzpicture}
}
{{\pmc} $\mathcal{P}_2$}
\end{center}
\end{minipage}}
\hspace{.005\textwidth}
\mbox{
\begin{minipage}[t]{.44\textwidth}
\begin{center}
{\tiny
\hspace{-0.4cm}
\begin{tikzpicture}[scale=1.4]

	\node[vertex, label={[label distance=-0.03cm]-45 :$\alpha$}](n1) at (1.2,1) {$s_1$};
    \node[vertex, label={[label distance=-0.03cm]-45 :$\alpha$}](n2) at (2.4,1) {$s_2$};
    \node(n3) at (3.6,1) {$\ldots$};
    \node(n5) at (3.15,0.4) {$\ldots$};
    \node[vertex, label={[label distance=-0.03cm]-45 :$\alpha$}](n4) at (4.4,1) {$s_n$};
	\node[vertex, label={[label distance=-0cm]+180 :$\beta$}] (n0) at (3,0)  {$s_0$};

	\draw[edge] (n1) to[out=45,in=45+90,min distance=4mm] node[above] {$p_1$} (n1);
	\draw[edge] (n1) to node[below left] {$q_1$} (n0);
    \draw[edge] (n1) to node[above] {$1{-}p_1{-}q_1$} (n2);
	\draw[edge] (n2) to[out=45,in=45+90,min distance=4mm] node[above] {$p_2$} (n2);
	\draw[edge] (n2) to node[below left] {$q_2$} (n0);
    \draw (n2) to node[above] {$1{-}p_2{-}q_2$} (n3);
    \draw[edge] (n3) to node[above] {} (n4);
	\draw[edge] (n4) to[out=45,in=45+90,min distance=4mm] node[above] {$p_n$} (n4);
	\draw[edge] (n4) to node[below right] {$1{-}p_n$} (n0);

	\draw[edge] (n0) to[out=-45,in=-45-90,min distance=4mm] node[below] {$1$} (n0);
\end{tikzpicture}
}
{{\pmc} $\mathcal{P}_n$}
\end{center}
\end{minipage}}
\caption{
	{\imc} $\mathcal{I}$ with three ${\pmc}$s 
    $\mathcal{P}_1$, $\mathcal{P}_2$, and $\mathcal{P}_n$
    entailed by $\mathcal{I}$
    \label{fig:imc_no_equiv_pmc}
    }%
\end{figure}
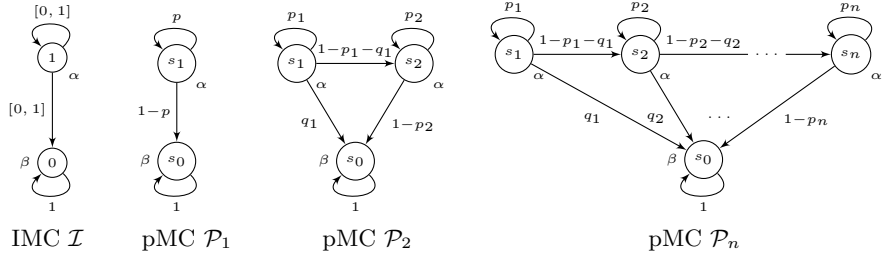

    \begin{enumerate}
    	\item Consider {\imc} $\mathcal{I}$ and {\pmc}s $\mathcal{P}_1$,
        	$\mathcal{P}_2$, and $\mathcal{P}_n$ (with $n \in \Nset$) from 	
            Figure~\ref{fig:imc_no_equiv_pmc}. 
        	presents an {\imc} $\mathcal{I}$ verifying case (1)
			$\mathcal{P}_1$, $\mathcal{P}_2$, and $\mathcal{P}_n$ (with $n \in \Nset$) 
            are entailed by $\mathcal{I}$ w.r.t. $\satisfactionImc$
            but none of them is equivalent to $\mathcal{I}$. 
            Indeed one needs an infinite countable number of states 
			for creating a ${\pmc}$ equivalent to $\mathcal{I}$ w.r.t. $\satisfactionImc$.
            However states space must be finite.
            
		\item Consider {\imc} $\mathcal{I^\prime}$ similar to $\mathcal{I}$ 
        	from Figure~\ref{fig:imc_no_equiv_pmc} excepted that 
        	the transition from $s_1$ to $s_0$ is replaced by the interval $[0.5, 1]$.
            Since the {\pimc} definition does not allow to bound values for parameters
            there is no equivalent $\mathcal{I^\prime}$ w.r.t. $\satisfactionImc$.
            
		\item Note that parameters in ${\pmc}$s enforce transitions 
        	in the concrete ${\mc}$s to receive the same value.
			Since parameters may range over continuous intervals there is no hope of modeling
			such set of Markov chains using a single $\imc$.
            Figure~\ref{fig:imc_no_equiv_pmc} illustrates this statement.
            
		\item Recall that $(\imc, \satisfactionImc)$ is more general than $(\imc, \satisfactionImcOnce)$. One cannot restrict the set of Markov chains satisfying an {\imc} w.r.t. $\satisfactionImc$ to be bisimilar to the set of Markov chain satisfying an {\imc} w.r.t. $\satisfactionImcOnce$.
            
        \item Same remark than item (3)
            
		\item Counter-example similar to which in statement (2) 
        (\ie in order to simulate the at-every-step satisfaction relation with the once-and-for-all satisfaction relation one needs an infinite uncountable number of states).
    \end{enumerate}

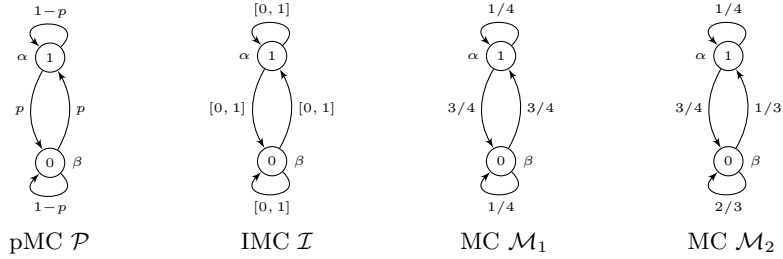
\begin{figure}[t]
\centering
\mbox{
\begin{minipage}[t]{.23\textwidth}
\centering
{\tiny
\hspace*{0.4cm}
\begin{tikzpicture}[scale=1.4]
	\node[vertex, label={[label distance=0cm]180 :$\alpha$}](n1) at (1.2,1) {$1$};
	\node[vertex, label={[label distance=0cm]0 :$\beta$}] (n0) at (1.2,0)  {$0$};

	\draw[edge] (n1) to[out=45,in=45+90,min distance=4mm] node[above] {$1{-}p$} (n1);
	\draw[edge, bend right] (n1) to node[left]  {$p$} (n0);
    \draw[edge, bend right] (n0) to node[right] {$p$} (n1);
	\draw[edge] (n0) to[out=-45,in=-45-90,min distance=4mm] node[below] {$1{-}p$} (n0);
\end{tikzpicture}
}
{{\newline}{\pmc} $\mathcal{P}$}
\end{minipage}}
\mbox{
\begin{minipage}[t]{.23\textwidth}
\centering
{\tiny
\hspace*{0.1cm}
\begin{tikzpicture}[scale=1.4]
	\node[vertex, label={[label distance=0cm]180 :$\alpha$}](n1) at (1.2,1) {$1$};
	\node[vertex, label={[label distance=0cm]0 :$\beta$}] (n0) at (1.2,0)  {$0$};

	\draw[edge] (n1) to[out=45,in=45+90,min distance=4mm] node[above] {$[0, 1]$} (n1);
	\draw[edge, bend right] (n1) to node[left]  {$[0, 1]$} (n0);
    \draw[edge, bend right] (n0) to node[right] {$[0, 1]$} (n1);
	\draw[edge] (n0) to[out=-45,in=-45-90,min distance=4mm] node[below] {$[0, 1]$} (n0);
\end{tikzpicture}
}
{{\newline}{\imc} $\mathcal{I}$}
\end{minipage}}
\mbox{
\begin{minipage}[t]{.23\textwidth}
\centering
{\tiny
\hspace*{0.2cm}
\begin{tikzpicture}[scale=1.4]
	\node[vertex, label={[label distance=0cm]180 :$\alpha$}](n1) at (1.2,1) {$1$};
	\node[vertex, label={[label distance=0cm]0 :$\beta$}] (n0) at (1.2,0)  {$0$};

	\draw[edge] (n1) to[out=45,in=45+90,min distance=4mm] node[above] {$1/4$} (n1);
	\draw[edge, bend right] (n1) to node[left]  {$3/4$} (n0);
    \draw[edge, bend right] (n0) to node[right] {$3/4$} (n1);
	\draw[edge] (n0) to[out=-45,in=-45-90,min distance=4mm] node[below] {$1/4$} (n0);
\end{tikzpicture}
}
{{\newline}{\mc} $\mathcal{M}_1$}
\end{minipage}}
\mbox{
\begin{minipage}[t]{.23\textwidth}
\begin{center}
{\tiny
\hspace*{0.2cm}
\begin{tikzpicture}[scale=1.4]
	\node[vertex, label={[label distance=0cm]180 :$\alpha$}](n1) at (1.2,1) {$1$};
	\node[vertex, label={[label distance=0cm]0 :$\beta$}] (n0) at (1.2,0)  {$0$};

	\draw[edge] (n1) to[out=45,in=45+90,min distance=4mm] node[above] {$1/4$} (n1);
	\draw[edge, bend right] (n1) to node[left]  {$3/4$} (n0);
    \draw[edge, bend right] (n0) to node[right] {$1/3$} (n1);
	\draw[edge] (n0) to[out=-45,in=-45-90,min distance=4mm] node[below] {$2/3$} (n0);
\end{tikzpicture}
}
{{\newline}{\mc} $\mathcal{M}_2$}
\end{center}
\end{minipage}}
\captionsetup{justification=centering}
\caption{
	{\pmc} $\mathcal{P}$, {\imc} $\mathcal{I}$,
    {\mc} $\mathcal{M}_1$, and {\mc} $\mathcal{M}_2$ s.t. \break
    $\mathcal{M}_1 \satisfactionPmc \mathcal{P}$ and $\mathcal{M}_1 \satisfactionImc \mathcal{I}$
    but
    $\mathcal{M}_2 \not\satisfactionPmc \mathcal{P}$ and $\mathcal{M}_2 \satisfactionImc \mathcal{I}$
    while $\mathcal{P}$ is entailed by $\mathcal{I}$
	\label{fig:pmc_with_no_eq_imc}
}%
\end{figure}

\begin{paragraph}{\textnormal{\textbf{Proposition~\ref{prop:sunccinctness_pimc_pmc_imc}.}}}
The Markov chain abstraction models can be ordered as follows w.r.t.
succinctness:
$({\pimcSet},\satisfactionPimcOnce) < (\pmcSet, \satisfactionPmc)$,
$({\pimcSet},\satisfactionPimcOnce) < (\imcSet, \satisfactionImcOnce)$ and
$({\pimcSet},\satisfactionPimc\nobreak) < (\imcSet, \satisfactionImc)$.
\end{paragraph}

\begin{proof}
Recall that the
{\pimcSet} model is a Markov chain abstraction model allowing to
declare parametric interval transitions, while the {\pmcSet} model
allows only parametric transitions (without intervals), and the
{\imcSet} model allows interval transitions without parameters.
Clearly, any {\pmc} and any {\imc} can be translated into a {\pimc}
with the right semantics (once-and-for-all for {\pmc}s and the chosen
{\imc} semantics for {\imc}s). This means that
$({\pimcSet},\satisfactionPimcOnce)$ is more succinct than {\pmcSet}
and {\pimcSet} is more succinct than {\imcSet} for both semantics.
Furthermore since {\pmcSet} and {\imcSet} are not comparable (cf Lemma~\ref{lem:compare_imc_and_pmc}), 
we have that the {\pimcSet}
abstraction model is strictly more succinct than the {\pmcSet}
abstraction model and than the {\imcSet} abstraction model with the
right semantics.\end{proof}

\section{Complements to Section~\ref{sec:qualitative-reachability}}
\subsection{{\csp} Encoding for Qualitative Reachability}

\begin{proof}[Proposition \ref{prop:csp_existential_consistency}]
Let $\mathcal{P} = (S,s_0,P,V,Y)$ be a $\pimc$.

$[\Rightarrow]$ The \csp\ $\Mec(\mathcal{P}) = (X,D,C)$ is satisfiable
implies that there exists a valuation $v$ of the variables in $X$ satisfying all the constraints in $C$.
Consider the \mc\ $\mathcal{M} = (S, s_0, p, V)$ such that 
$p(s, s^\prime) = v(\transition{s}{s^\prime})$, for all $\transition{s}{s^\prime} \in \transitionSet$
and $p(s, s^\prime) = 0$ otherwise.

Firstly, we show by induction that for any state $s$ in $S$:
``if $s$ is reachable in $\mathcal{M}$ then $v(\rho_s)$ equals to {\true}''.
This is correct for the initial state $s_0$ thanks to the Constraint (1).
Let $s$ be a state in $S$ and assume 
that the property is correct for all its predecessors.
By the Constraints (2), $v(\rho_s)$ equals to {\true} if 
there exists at least one predecessor $s^{\prime\prime} \neq s$ reaching $s$ with a non-zero probability 
(\ie $v(\transition{s^{\prime\prime}}{s}) \neq 0$).
This is only possible by the Constraint (3) if $v(\rho_{s^{\prime\prime}})$ equals to {\true}.
Thus $v(\rho_s)$ equals to {\true} if there exists one reachable state $s^{\prime\prime}$ s.t. 
$v(\transition{s^{\prime\prime}}{s}) \neq 0$.

Secondly, we show that $\mathcal{M}$ satisfies the \pimc\ $\mathcal{P}$.
We use Theorem~4 from \cite{DelahayeLP16} stating that
$\satisfactionPimc$ and $\satisfactionPimcOnce$ are equivalent w.r.t. 
qualitative reachability.
We proved above that for all reachable states $s$ in $\mathcal{M}$,
we have $v(\rho_s)$ equals to {\true}.
By the Constraints (5) it implies that 
that for all reachable states $s$ in $\mathcal{M}$: $p(s,s^\prime) \in P(s,s^\prime)$ for all $s$ and $s^\prime$.%
\footnote{As illustrated in Example \ref{ex:model_consistency},
$\mathcal{M}$ is not a well formed {\mc} since some unreachable states do not respect the probability distribution property. 
However, one can correct it by simply setting one of its outgoing transition to $1$ for each unreachable state.}

$[\Leftarrow]$ The \pimc\ $\mathcal{P}$ is consistent implies by the Theorem 4 from \cite{DelahayeLP16} stating that
$\satisfactionPimc$ and $\satisfactionPimcOnce$ are equivalent w.r.t. 
qualitative reachability,
that there exists an implementation of the form $\mathcal{M} = (S, s_0, p, V )$ where, 
for all reachable states $s$ in $\mathcal{M}$, 
it holds that $p(s,s^\prime) \in P(s,s^\prime)$ for all $s^\prime$ in $S$.
Consider $\mathcal{M}^\prime = (S, s_0, p^\prime, V )$ s.t. 
for each non reachable state $s$ in $S$: $p^\prime(s, s^\prime) = 0$, for all $s^\prime \in S$.
The valuation $v$ is s.t. $v(\rho_s)$ equals {\true} iff $s$ is reachable in $\mathcal{M}$,
$v(\transition{s}{s^{\prime}}) = p^\prime(s,s^\prime)$,
and for each parameter $y \in Y$ a valid value can be selected according to $p$ and $P$ when considering reachable states.
Finally, by construction, $v$ satisfies the \csp\ $\Mec(\mathcal{P})$.\qed
\end{proof}

\section{Complements to Section~\ref{sec:quantitative}}
\subsection{Equivalence of $\satisfactionImcOnce$ and
  $\satisfactionImc$ w.r.t quantitative reachability}\label{ap:equiv_imc_semantics}

We first introduce some notations.
Let $\mathcal{I} = (S,s_0,P,V^I)$ be an {\imc} and
$\mathcal{M}=(T,t_0,p,V^M)$ be an {\mc} s.t. $\mathcal{M} \satisfactionImc \mathcal{I}$.
Let $\mathcal{R} \subseteq T \times S$ be a satisfaction relation between $\mathcal{M}$ and $\mathcal{I}$.
For all $t \in T$ we write $\mathcal{R}(t)$ for the set $\{s \in S \mid t~\mathcal{R}~s \}$,
and for all $s \in S$ we write $\mathcal{R}^{-1}(t)$ for the set $\{t \in T \mid s~\mathcal{R}~t \}$.
Furthermore we say that $\mathcal{M}$ satisfies $\mathcal{I}$ with degree $n$ 
(written \nsatisfaction{$n$}{\mathcal{M}}{\mathcal{I}}) 
if $\mathcal{M}$ satisfies $\mathcal{I}$
with a satisfaction relation $\mathcal{R}$
s.t. each state $t \in T$ is associated by $\mathcal{R}$ to at most $n$ states from $S$
(\ie $|\mathcal{R}(t)| \leq n$);
	$\mathcal{M}$ satisfies $\mathcal{I}$ with the same structure than $\mathcal{I}$
	if $\mathcal{M}$ satisfies $\mathcal{I}$
	with a satisfaction relation $\mathcal{R}$
	s.t. each state $t \in T$ is associated to at most one state from $S$
    and each state $s \in S$ is associated to at most one state from $T$
    (\ie $|\mathcal{R}(t)| \leq 1$ for all $t \in T$ and 
    	$|\mathcal{R}^{-1}(s)| \leq 1$ for all $s \in S$).

\begin{proposition}\label{prop:nsat_to_bisimilar_1sat}
	Let $\mathcal{I}$ be an \imc.
	If a {\mc} $\mathcal{M}$ satisfies $\mathcal{I}$ with degree $n \in \Nset$ 
	then there exists a {\mc} $\mathcal{M^\prime}$ satisfying $\mathcal{I}$ with degree $1$ 
	such that $\mathcal{M}$ and $\mathcal{M}^\prime$ are bisimilar.
\end{proposition}

The main idea for proving Proposition \ref{prop:nsat_to_bisimilar_1sat}
is that if an ${\mc}$ $\mathcal{M}$ with states space $T$ 
satisfies an ${\imc}$ $\mathcal{I}$ with a states space $S$
according to a satisfaction relation $\mathcal{R}$
then, each state $t$ related by $\mathcal{R}$ to many states $s_1, \ldots, s_n$ (with $n > 1$)
can be split in $n$ states $t^1, \ldots, t^n$.
The derived ${\mc}$ will satisfy $\mathcal{I}$ with a satisfaction relation $\mathcal{R}^\prime$
where each $t_i$ is only associated by $\mathcal{R}^\prime$ to the state $s_i$ ($i \leq n$).
This $\mathcal{M^\prime}$ will be bisimilar to $\mathcal{M}$ and 
it will satisfy $\mathcal{I}$ with degree $1$.
Note that by construction the size of the resulting ${\mc}$ is in $O(|\mathcal{M}|\times|\mathcal{I}|)$.

\begin{proof}[Proposition \ref{prop:nsat_to_bisimilar_1sat}]
	Let $\mathcal{I} = (S,s_0,P,V^I)$ be an {\imc} and
	$\mathcal{M}=(T,t_0,p,V)$ be a \mc.
	If $\mathcal{M}$ satisfies $\mathcal{I}$ (with degree $n$)
	then there exists a satisfaction relation $\mathcal{R}$ verifying 
    the $\satisfactionImc$ satisfaction relation.	
	For each association $t~\mathcal{R}~s$,
	we write $\delta_t^s$ the correspondence function chosen for this pair of states.
	$\mathcal{M}$ satisfies $\mathcal{I}$ with degree $n$ 
	means that each state in $\mathcal{M}$ is associated by $\mathcal{R}$ to at most $n$ states in $\mathcal{I}$.
	To construct a {\mc} $\mathcal{M^\prime}$ satisfying $\mathcal{I}$ with degree $1$
	we create one state in $\mathcal{M^\prime}$ 
	per association $(t,s)$ in $\mathcal{R}$.
	Formally, let $\mathcal{M^\prime}$ be equal to $(U, u_0, p^\prime, V^\prime)$ such that 
	$U = \{u_t^s \mid t~\mathcal{R}~s\}$, 
	$u_0 = u_{t_0}^{s_0}$, 
	$V^\prime = \{(u_t^s, v) \mid v = V(t)\}$, and
	$p^\prime(u_t^s)(u_{t^\prime}^{s^\prime}) = p(t)(t^\prime) \times \delta_t^s(t^\prime)(s^\prime)$.
	Following computation shows that the outgoing probabilities given by $p^\prime$ form a probability distribution
	for each state in $\mathcal{M^\prime}$
	and thus that $\mathcal{M^\prime}$ is a {\mc}.
	\par\noindent
	\begin{align*}
		&
		\ssum{t^\prime \mathcal{R} s^\prime}{} p^\prime(u_t^s)(u_{t^\prime}^{s^\prime})
		=
		\ssum{t^\prime \mathcal{R} s^\prime}{}  p(t)(t^\prime) \times \delta_t^s(t^\prime)(s^\prime) \\
		= &
		\ssum{t^\prime \in T}{}  p(t)(t^\prime) \times \ssum{s^\prime \in S}{} \delta_t^s(t^\prime)(s^\prime)
		=
		\ssum{t^\prime \in T}{}  p(t)(t^\prime) \times 1	
		=
		1
	\end{align*}
	
	Finally, by construction of $\mathcal{M^\prime}$ based on $\mathcal{M}$
	which satisfies $\mathcal{I}$,
	we get that $\mathcal{R^\prime} = \{(u_t^s,s) \mid t \in T, s \in S \}$
	is a satisfaction relation between $\mathcal{M^\prime}$ and $\mathcal{I}$.
	Furthermore $|\{s \mid u~\mathcal{R^\prime}~s\}|$ equals at most one.
	Thus, we get that $\mathcal{M^\prime}$ satisfies $\mathcal{I}$ with degree $1$.
	
	Consider the relation $\mathcal{B}^\prime = \{ (u_t^s, t) \subseteq U \times T \mid t~\mathcal{R}~s \}$.
	We note $\mathcal{B}$ the closure of $\mathcal{B}^\prime$ by transitivity, reflexivity and symetry
	(\ie $\mathcal{B}$ is the minimal equivalence relation based on $\mathcal{B}^\prime$).
	We prove that $\mathcal{B}$ is a bisimulation relation between $\mathcal{M}$ and $\mathcal{M^\prime}$.
	By construction each equivalence class from $\mathcal{B}$
	contains exactly one state $t$ from $T$ and
	all the states $u_t^s$ such that $t~\mathcal{R}~s$.
	Let $(u_t^s, t)$ be in $\mathcal{B}$,
	$t^\prime$ be a state in $T$,
	and $B$ be the equivalence class from $\mathcal{B}$ containing $t^\prime$
	(\ie $B$ is the set $\{ t^\prime \} \cup \{ u_{t^\prime}^{s^\prime} \in U \mid s^{\prime} \in S \text{ and } t^{\prime}~\mathcal{R}~s^{\prime} \}$).
	Firstly note that by construction the labels agree on $u_t^s$ and $t$: $V^\prime(u_t^s) = V(t)$.
	Secondly the following computation shows that $p^\prime(u_t^s)(B \cap U)$ equals to $p(t)(B \cap T)$
	and thus that $u_t^s$ and $t$ are bisimilar:
	\par\noindent
	\begin{align*}
		p^\prime(u_t^s)(B \cap U) &
		=
		\ssum{\hspace*{0.6cm}u_{t^\prime}^{s^\prime} \in B \cap U}{} p^\prime(u_t^s)(u_{t^\prime}^{s^\prime})
		=
		\ssum{\hspace*{0.6cm}u_{t^\prime}^{s^\prime} \in B \cap U}{} p(t)(t^\prime) \times \delta_t^s(t^\prime)(s^\prime) \\
		& 
		=
		\ssum{\hspace*{1cm}\{s^\prime \in S \mid s^\prime \mathcal{R} t^\prime \}}{} p(t)(t^\prime) \times \delta_t^s(t^\prime)(s^\prime)
		=
		p(t)(t^\prime) \times \ssum{\hspace*{1cm}\{s^\prime \in S \mid s^\prime \mathcal{R} t^\prime \}}{} \delta_t^s(t^\prime)(s^\prime) \\
		&
		= 		
		p(t)(t^\prime) \times 1	
		=
		p(t)(\{t^\prime\}) 
		=
		p(t)(B \cap T)
	\end{align*}
	~
	\qed
\end{proof}

\begin{corollary}\label{lem:1sat_same_proba}
	Let $\mathcal{I}$ be an \imc,
	$\mathcal{M}$ be a {\mc} satisfying $\mathcal{I}$, and
	$\gamma$ be a {\pctlStar} formulae.
	There exists a {\mc} $\mathcal{M^\prime}$ satisfying $\mathcal{I}$ with degree $1$
	such that the probability $\Proba^{\mathcal{M}^\prime}(\gamma)$ equals the probability $\Proba^{\mathcal{M}}(\gamma)$.
\end{corollary}

Corollary~\ref{lem:1sat_same_proba} is derived from Proposition \ref{prop:nsat_to_bisimilar_1sat}
joined with the probability preservation of the PCTL* formulae on bisimilar Markov chains
(see~\cite{Baier2008PMC}, Theorem~10.67, p.813).
Corollary~\ref{lem:1sat_same_proba} 
allows to reduce to one the number of states in the {\pimc} $\mathcal{I}$
satisfied by each state in the {\mc} $\mathcal{M}$ while preserving probabilities.

\begin{lemma}\label{lem:reach_with_less_states}
	Let $\mathcal{I} = (S,s_0,P,V)$ be an {\imc},
	$\mathcal{M} = (T,t_0,p,V)$ be a {\mc} satisfying $\mathcal{I}$ with degree $1$, and
	$\alpha \subseteq A$ be a proposition.
	If $\mathcal{M}$ does not have the same structure than $\mathcal{I}$
	then there exists an {\mc} $\mathcal{M}_1$ (resp. $\mathcal{M}_2$) 
	satisfying $\mathcal{I}$ with a set of states $S_1$ (resp. $S_2$)
	s.t. $S_1 \subset S$ and $\Proba^{\mathcal{M}_1}(\ltlExists \alpha) \leq \Proba^{\mathcal{M}}(\ltlExists \alpha)$
	(resp. $S_2 \subset S$ and $\Proba^{\mathcal{M}_2}(\ltlExists \alpha) \geq \Proba^{\mathcal{M}}(\ltlExists \alpha)$). 
\end{lemma}

Lemma~\ref{lem:reach_with_less_states} reduces the number of states in $\mathcal{M}$ while preserving the maximal or minimal reachability probability.
This lemma has a constructive proof.
Here is the main idea of the proof.
We first select one state $s$ from the {\imc} $\mathcal{I}$
which is satisfied by many states $t_1, \ldots, t_n$ in $\mathcal{M}$.
The ${\mc}$ $\mathcal{M}^\prime$ keeping the state $t_k$ minimizing the probability of reaching $\alpha$ in $\mathcal{M}$ and removing all the other states $t_i$
(\ie remove the states $t_i$ s.t. $i \neq k$ and 
move the transitions arriving to a state $t_i$ s.t. $i \neq k$ to arrive to the state $t_k$)
will have less states than $\mathcal{M}$ and verifies $\Proba^{\mathcal{M}_1}(\ltlExists \alpha) \leq \Proba^{\mathcal{M}}(\ltlExists \alpha)$.

\begin{proof}[Lemma \ref{lem:reach_with_less_states}]
	Let $\mathcal{I} = (S,s_0,P,V^I)$ be an {\imc} and
	$\mathcal{M} = (T,t_0,p,V)$ be a {\mc} satisfying $\mathcal{I}$ with degree $1$.
	We write $\mathcal{R}$ the satisfaction relation between $\mathcal{M}$ and $\mathcal{I}$ with degree $1$.
	The following proves in 3 steps the $\Proba^{\mathcal{M}_1}(\ltlExists \alpha) \leq \Proba^{\mathcal{M}}(\ltlExists \alpha)$ case.
		
	\begin{enumerate}
	\item
	We would like to construct a {\mc} $\mathcal{M}^\prime$ satisfying $\mathcal{I}$ 
	with less states than $\mathcal{M}^\prime$
	such that $\Proba^{\mathcal{M}^\prime}(\ltlExists \alpha) \leq \Proba^{\mathcal{M}}(\ltlExists \alpha)$.
	Since the degree of $\mathcal{R}$ equals to $1$ each state $t$ in $T$ is associated to at most one state $s$ in $S$.
	Furthermore, since $\mathcal{M}$ does not have the same structure than $\mathcal{I}$
	then there exists at most one state from $S$ which is associated by $\mathcal{R}$ to many states from $T$.
	Let 
	$\bar{s}$ be a state from $S$ such that $|\mathcal{R}^{-1}(s)| \geq 2$,
	$\bar{T} = \{t_1, \ldots, t_n\}$ be the set $\mathcal{R}^{-1}(s)$
	where the $t_i$ are ordered by decreasing probability of reaching~$\alpha$
	(\ie
		$\Proba^{\mathcal{M}}_{t_i}(\ltlExists \alpha) \geq \Proba^{\mathcal{M}}_{t_{i+1}}(\ltlExists \alpha)$
		for all $1 \leq i < n$).
	In the following we refer $\bar{t}$ as $t_n$.
	We produce $\mathcal{M}^\prime$ from $\mathcal{M}$ 
	by replacing all the transitions going to a state $t_1, \ldots, t_{n-1}$
	by a transition going to $t_{n}$,
	and by removing the corresponding states.
	Formally $\mathcal{M}^\prime = (T^\prime,t_0,p^\prime,V^\prime)$ s.t. 
	$T^\prime = (T \setminus \bar{T}) \cup \{\bar{t}\}$,
	$V^\prime$ is the restriction of $V$ on $T^\prime$, and
	for all $t,t^\prime \in T^\prime$:
	$p^\prime(t)(t^\prime) = p(t)(t^\prime)$ if $t^\prime \neq \bar{t}$ and 
	$p^\prime(t)(t^\prime) = \ssum{t^\prime \in \bar{T}}{} p(t)(t^\prime)$ otherwise.

	\par\noindent
	\begin{align*}
		\ssum{t^\prime \in T^\prime}{} p^\prime(t)(t^\prime)
		=
		&
		\ssum{\hspace{0.7cm}t^\prime \in T^\prime \setminus \{\bar{t}\}}{} p^\prime(t)(t^\prime)
		\hspace{0.1cm} + \hspace{0.1cm} 
		p^\prime(t)(\bar{t})
\\
		\stackrel{(1)}{=} &
		\ssum{\hspace{0.7cm}t^\prime \in T^\prime \setminus \{\bar{t}\}}{} p(t)(t^\prime)
		\hspace{0.1cm} + \hspace{0.1cm} 
		\ssum{\hspace{0cm}t^\prime \in \bar{T}}{} p(t)(t^\prime)
\\
		= &
		\ssum{\hspace{1cm}t^\prime \in T^\prime \setminus \{\bar{t}\} \cup \bar{T}}{} p(t)(t^\prime)
		\hspace{0.1cm} \stackrel{(2)}{=} \hspace{0.1cm} 	
		\ssum{\hspace{0cm}t^\prime \in T}{} p(t)(t^\prime)
		\hspace{0.1cm} = \hspace{0.1cm} 
		1
	\end{align*}

	Previous computation holds for each state $t$ in $\mathcal{M^\prime}$.
	It shows that the outgoing probabilities given by $p^\prime$ form a probability distribution
	for each state in $\mathcal{M^\prime}$
	and thus that $\mathcal{M^\prime}$ is a {\mc}.
	Note that step ($1$) comes from the definition of $p^\prime$ with respect to $p$ and 
	that step ($2$) comes from the definition of $T^\prime$ according to $\bar{T}$ and $\bar{t}$.
	
	\item 
	We now prove that $\mathcal{M}^\prime$ implements $\mathcal{I}$.
	$\mathcal{M}$ satisfies $\mathcal{I}$ implies that there a exists a satisfaction relation $\mathcal{R}$ between $\mathcal{M}$ and $\mathcal{I}$.
	Let $\mathcal{R}^\prime \subseteq T \times S$ be s.t. 
	$t~\mathcal{R}^\prime~s$ if $t~\mathcal{R}~s$
	and  $\bar{t}~\mathcal{R}^\prime~\bar{s}$ if there exists a state 
    $t^\prime \in \bar{T}$ s.t. $t^\prime~\mathcal{R}~\bar{s}$.
	We prove that $\mathcal{R}^\prime$ is a satisfaction relation between $\mathcal{M}^\prime$ and $\mathcal{I}$.
	For each pair $(t,s) \in \mathcal{R}$ we note $\delta_{(s,t)}$ the correspondance function given by the satisfaction relation $\mathcal{R}$. 
	Let $(t,s)$ be in $\mathcal{R}^\prime$ and
	$\delta^\prime : T^\prime \to (S \to [0, 1])$ be s.t.
	$\delta^\prime(t^\prime)(s^\prime) = \delta_{(t,s)}(t^\prime)(s^\prime)$ if $t^\prime \neq \bar{t}$ and 
	$\delta^\prime(t^\prime)(s^\prime) = max_{t^\prime \in \bar{T}}(\delta_{(t,s)}(t^\prime)(s^\prime))$ otherwise.
	$\delta^\prime$ is a correspondence function for the pair $(t,s)$ in $\mathcal{R}^\prime$ such as required by the {\satisfactionImc} satisfaction relation:
	\begin{enumerate}
		\item Let $t^\prime$ be in $T$. If $t^\prime \neq \bar{t}$ 
		then $\delta^\prime(t^\prime)$ is equivalent to $\delta_{(t,s)}(t^\prime)(s^\prime)$
		which is by definition a distribution on $S$. 
		Otherwise $t^\prime = \bar{t}$ and the following computation proves 
		that $\delta^\prime(\bar{t})$ is a distribution on $S$.
		For the step (1) remind 
		that $\mathcal{R}$ is a satisfaction relation with degree $1$ and 
		that $\bar{t}~\mathcal{R}~\bar{s}$.
		This implies that $\delta_{(t,s)}(\bar{t})(s^\prime)$ equals to zero for all $s^\prime \neq \bar{s}$.
		For the step (2), $\mathcal{R}$ is a satisfaction relation with degree $1$ implies 
		that $\delta_{(t,s)}(t^{\prime})(s^\prime)$ equals to $0$ or $1$ for all $t^{\prime} \in T$ and $s^\prime \in S$.
		Finally the recursive definition of the satisfaction relation $\mathcal{R}$
		implies that there exists at least one state $t^{\prime\prime} \in \bar{T}$ s.t. $\delta_{(t,s)}(t^{\prime\prime})(\bar{s})$ does not equal to zero
		(\ie equals to one).
		
	\par\noindent
	\begin{align*}
		\ssum{s^\prime \in S}{} \delta^\prime(\bar{t})(s^\prime)
		=
		&
		\ssum{\hspace{0.5cm}s^\prime \in S \setminus \{\bar{s}\}}{} \delta^\prime(\bar{t})(s^\prime)
		\hspace{0.1cm} + \hspace{0.1cm} 
		\delta^\prime(\bar{t})(\bar{s})
\\
		= &
		\ssum{\hspace{0.5cm}s^\prime \in S \setminus \{\bar{s}\}}{} \delta_{(t,s)}(\bar{t})(s^\prime)
		\hspace{0.1cm} + \hspace{0.1cm} 
		max_{t^{\prime\prime} \in \bar{T}}(\delta_{(t,s)}(t^{\prime\prime})(\bar{s}))
\\
		\stackrel{(1)}{=} &~
		max_{t^{\prime\prime} \in \bar{T}}(\delta_{(t,s)}(t^{\prime\prime})(s^\prime))
\\
		\stackrel{(2)}{=} &~
		1
	\end{align*}
			
		\item
			Let $s^\prime$ be in $S$. 
			Step (1) uses the definition of $p^\prime$ according to $p$.
			Step (2) uses the definition of $\delta^\prime$ according to $\delta_{(t,s)}$.
			Step (3) comes from the fact that for all $t, t^\prime \in T \times \bar{T}$, 
			we have by the definition of the satisfaction relation $\mathcal{R}$ with degree $1$
			and by construction of $\bar{T}$
			that if $p(t,t^\prime) \neq 0$ then $\delta_{(t)(s)}(t^\prime, \bar{s}) = 1$ and $\delta_{(t,s)}(t^\prime)(s^\prime) = 0$ for all $s^\prime \neq \bar{s}$.
			Finally, step (4) comes from the definition of the correspondence function $\delta_{(t,s)}$ for the pair $(t,s)$ in $\mathcal{R}$.
	\par\noindent
	\allowdisplaybreaks
	\begin{align*}
		& \ssum{t^\prime \in T^\prime}{} p^\prime(t)(t^\prime) \times \delta^\prime(t^\prime)(s^\prime)
\\
		=
		&
		\ssum{\hspace{0.7cm}t^\prime \in T^\prime \setminus \{\bar{t}\}}{} p^\prime(t)(t^\prime) \times \delta^\prime(t^\prime)(s^\prime)
		\hspace{0.1cm} + \hspace{0.1cm} 
		p^\prime(t, \bar{t}) \times \delta^\prime(\bar{t})(s^\prime)
\\
		\stackrel{(1)}{=} &
		\ssum{\hspace{0.7cm}t^\prime \in T^\prime \setminus \{\bar{t}\}}{} p(t)(t^\prime) \times \delta^\prime(t^\prime)(s^\prime)
		\hspace{0.1cm} + \hspace{0.1cm} 
		\ssum{\hspace{0cm}t^\prime \in \bar{T}}{} p(t)(t^\prime) \times \delta^\prime(\bar{t})(s^\prime)
\\
		\stackrel{(2)}{=} &
		\ssum{\hspace{0.7cm}t^\prime \in T^\prime \setminus \{\bar{t}\}}{} p(t)(t^\prime) \times \delta_{(t,s)}(t^\prime)(s^\prime)
		\hspace{0.1cm} + \hspace{0.1cm} 
		\ssum{\hspace{0cm}t^\prime \in \bar{T}}{} p(t)(t^\prime) \times max_{t^{\prime\prime} \in \bar{T}}(\delta_{(t,s)}(t^{\prime\prime})(s^\prime))
\\
		\stackrel{(3)}{=} &
		\ssum{\hspace{0.7cm}t^\prime \in T^\prime \setminus \{\bar{t}\}}{} p(t)(t^\prime) \times \delta_{(t,s)}(t^\prime)(s^\prime)
		\hspace{0.1cm} + \hspace{0.1cm} 
		\ssum{\hspace{0cm}t^\prime \in \bar{T}}{} p(t)(t^\prime) \times  \delta_{(t,s)}(t^{\prime})(s^\prime)
\\
		= &
		\ssum{\hspace{1cm}t^\prime \in T^\prime \setminus \{\bar{t}\} \cup \bar{T}}{} p(t)(t^\prime) \times \delta_{(t,s)}(t^\prime)(s^\prime)
		\hspace{0.1cm} {=} \hspace{0.1cm} 	
		\ssum{\hspace{0cm}t^\prime \in T}{} p(t)(t^\prime) \times \delta_{(t,s)}(t^\prime)(s^\prime)
\\
		\stackrel{(4)}\in & \hspace{0.1cm} 
		P(s, s^\prime)
	\end{align*}
	
		\item Let $t^\prime$ be in $T^\prime$ and $s^\prime$ be in $S$. 
		We have by construction of $\mathcal{R}^\prime$ from $\mathcal{R}$
		that if $\delta^\prime(t^\prime)(s^\prime) > 0$ then $(t^\prime, s^\prime) \in \mathcal{R}$.
	
	\end{enumerate}
	
	\item Ne now prove that the probability of reaching $\alpha$ from $\bar{t}$
	is lower in $\mathcal{M^\prime}$ than in $\mathcal{M}$.
	We consider the {\mc} $\mathcal{M}^{\prime\prime}$ from $\mathcal{M}$ 
	where the states containing the label $\alpha$ are replaced by absorbing states.
	Formally $\mathcal{M}^{\prime\prime} = (T,t_0,p^{\prime\prime},V)$ such that
	for all $t,t^\prime \in T$:
	$p^{\prime\prime}(t, t^\prime) = p(t, t^\prime)$ if $\alpha \not\subseteq V(t)$
	else $p^{\prime\prime}(t, t^\prime) = 1$ if $t=t^\prime$ and $p^{\prime\prime}(t, t^\prime) = 0$ otherwise.
	By definition of the reachability property we get that
	$\Proba^{\mathcal{M}^{\prime\prime}}_t(\ltlExists \alpha)$ equals to $\Proba^{\mathcal{M}}_t(\ltlExists \alpha)$
	for all state $t$ in $T^\prime$.
	Following computation concludes the proof.
	Step (1) comes from Lemma~\ref{lem:proba_with_loops}.
    Step (2) comes by construction of $\mathcal{M}^\prime$ from $\mathcal{M}$.
	Step (3) comes by construction of $\mathcal{M}^{\prime\prime}$ from $\mathcal{M}$
    where states labeled with $\alpha$ are absorbing states.
    Step (4) comes from the fact that 
    	$\Proba^{\mathcal{M}^{\prime\prime}}_{t_n}(\ltlExists \alpha)$
        is equal to 
        $\Proba^{\mathcal{M}^{\prime\prime}}_{t_n}(\neg(t_1 \lor \ldots \lor t_n) \ltlUntil \alpha) + 
		    \Sigma_{1 \leq i \leq n}
            	\Proba^{\mathcal{M}^{\prime\prime}}_{t_n}(
                	\neg(t_1 \lor \ldots \lor t_n) \ltlUntil t_{i})
                \times \Proba^{\mathcal{M}^{\prime\prime}}_{t_i}(\ltlExists \alpha)$.
	Step (5) uses the fact that $\Proba^{\mathcal{M}}_{t_i}(\ltlExists \alpha) \geq \Proba^{\mathcal{M}}_{t_{n}}(\ltlExists \alpha)$ for all $1 \leq i \leq n$
    and by construction this is also correct in $\mathcal{M}^{\prime\prime}$.
    Last steps are straightforward.
	
	\par\noindent
	\allowdisplaybreaks
	\begin{align*}
    	&
		\Proba^{\mathcal{M}^{\prime}}_{\bar{t}}(\ltlExists \alpha)
        \\
	 	\stackrel{(1)}{=} 
		&
        \frac{
			\Proba^{\mathcal{M}^{\prime}}_{\bar{t}}(\neg \bar{t} \ltlUntil \alpha)
		}{
		    1 - \Proba^{\mathcal{M}^{\prime}}_{\bar{t}}(\neg\alpha \ltlUntil \bar{t})
		} 
        \\
	 	\stackrel{(2)}{=} 
		&
        \frac{
			\Proba^{\mathcal{M}}_{t_n}(\neg(t_1 \lor \ldots \lor t_n) \ltlUntil \alpha)
		}{
		    1 - \Proba^{\mathcal{M}}_{t_n}(\neg\alpha \ltlUntil (t_1 \lor \ldots \lor t_n))
		} 
        \\
		\stackrel{(3)}{=} 
		&
        \frac{
			\Proba^{\mathcal{M}^{\prime\prime}}_{t_n}(\neg(t_1 \lor \ldots \lor t_n) \ltlUntil \alpha)
		}{
		    1 - \Proba^{\mathcal{M}^{\prime\prime}}_{t_n}(\ltlExists (t_1 \lor \ldots \lor t_n))
		} 
        \\
	 	\stackrel{(4)}{=} 
		&
		\frac{
			\Proba^{\mathcal{M}^{\prime\prime}}_{t_n}(\ltlExists \alpha) - 
		    \ssum{1 \leq i \leq n}{} 
            	\Proba^{\mathcal{M}^{\prime\prime}}_{t_n}(
                	\neg(t_1 \lor \ldots \lor t_n) \ltlUntil t_{i})
                \times \Proba^{\mathcal{M}^{\prime\prime}}_{t_i}(\ltlExists \alpha)
		}{
		    1 - \Proba^{\mathcal{M}^{\prime\prime}}_{t_n}(\ltlExists (t_1 \lor \ldots \lor t_n))
		} 
        \\
	 	\stackrel{(5)}{\leq} 
		&
		\frac{
			\Proba^{\mathcal{M}^{\prime\prime}}_{t_n}(\ltlExists \alpha) - 
		    \ssum{1 \leq i \leq n}{} 
            	\Proba^{\mathcal{M}^{\prime\prime}}_{t_n}(
                	\neg(t_1 \lor \ldots \lor t_n) \ltlUntil t_{i})
                \times \Proba^{\mathcal{M}^{\prime\prime}}_{t_n}(\ltlExists \alpha)
		}{
		    1 - \Proba^{\mathcal{M}^{\prime\prime}}_{t_n}(\ltlExists (t_1 \lor \ldots \lor t_n))
		} 
        \\
	 	\stackrel{(6)}{=} 
		&
		\frac{
			\Proba^{\mathcal{M}^{\prime\prime}}_{t_n}(\ltlExists \alpha) \times
            (1 - 
		    \ssum{1 \leq i \leq n}{} 
            	\Proba^{\mathcal{M}^{\prime\prime}}_{t_n}(
                	\neg(t_1 \lor \ldots \lor t_n) \ltlUntil t_{i})
            )
		}{
		    1 - \Proba^{\mathcal{M}^{\prime\prime}}_{t_n}(\ltlExists (t_1 \lor \ldots \lor t_n))
		} 
        \\
	 	\stackrel{(7)}{=} 
		&
		\frac{
			\Proba^{\mathcal{M}^{\prime\prime}}_{t_n}(\ltlExists \alpha) \times
            (1 - 
		    \Proba^{\mathcal{M}^{\prime\prime}}_{t_n}(\ltlExists (t_1 \lor \ldots \lor t_n))
            )
		}{
		    1 - \Proba^{\mathcal{M}^{\prime\prime}}_{t_n}(\ltlExists (t_1 \lor \ldots \lor t_n))
		} 
        \\
		= 
		& ~
		\Proba^{\mathcal{M}^{\prime\prime}}_{t_n}(\ltlExists \alpha) \\
		= 
		& ~ \Proba^{\mathcal{M}}_{\bar{t}}(\ltlExists \alpha)
	\end{align*}

	\end{enumerate}
	
	The same method can be used for proving the $\Proba^{\mathcal{M}_2}(\ltlExists \alpha) \geq \Proba^{\mathcal{M}}(\ltlExists \alpha)$ case
	by defining $\bar{T} = \{t_1, \ldots, t_n\}$ be the set $\mathcal{R}^{-1}(s)$ s.t.
	the $t_i$ are ordered by {\em increasing} probability of reaching~$\alpha$.
	Thereby the $\leq$ symbol at step (5) for the computation of $\Proba^{\mathcal{M}^{\prime}}_{\bar{t}}(\ltlExists \alpha)$
	is replaced by the $\geq$ symbol.\qed
\end{proof}

\begin{lemma}\label{lem:proba_with_loops}
	Let $\mathcal{M} = (S, s_0 , p, V)$ be a {\mc},
	$\alpha \subseteq A$ be a proposition,
	and $s$ be a state from $S$.
    Then
    $$\Proba^\mathcal{M}_s(\ltlExists \alpha) =
    \frac{
    	\Proba^\mathcal{M}_s(\neg s \ltlUntil \alpha)}{
    	1 - \Proba^\mathcal{M}_s(\neg\alpha \ltlUntil s)}$$
\end{lemma}

Lemma~\ref{lem:proba_with_loops} is called in the proof of Lemma~\ref{lem:reach_with_less_states}.

\begin{proof}[Lemma \ref{lem:proba_with_loops}]
	Let $S^\prime$ be the subset of $S$ containing
    all the states labeled with $\alpha$ in $\mathcal{M}$.
	We write $\Omega_n$ with $n \in \Nset^*$ the set containing all the paths $\omega$ starting from $s$ s.t. state $s$ appears exactly $n$ times in $\omega$ and no state in $\omega$ is labeled with $\alpha$.
    Formally $\Omega_n$ contains all the 
    $\omega = s_1, \ldots, s_k \in S^k$
    s.t. $k \in \Nset$, $s_1$ is equal to $s$, 
    $|\{ i \in [1,k] \mid s_i = s \}| = n$, and
    $\alpha \not\subseteq V(s_i)$ for all $i \in [1,k]$.
    We get by (a) that $(\Proba^\mathcal{M}_s(\Omega_n \times S^\prime))_{n \geq 1}$ is a geometric series.
	In (b) we partition the paths reaching $\alpha$
    according to the $\Omega_n$ sets 
    and we use the geometric series of the probabilities to retrieve the required result.
    
	\begin{align*}
    	\text{(a)~} \Proba^\mathcal{M}_s(\Omega_n \times S^\prime) &
        = \Proba^\mathcal{M}_s(\Omega_1 \times \Omega_{n-1} \times S^\prime)
        \\
        & = 
        \Proba^\mathcal{M}_s(\Omega_1 \times \{ s \}) \times
		\Proba^\mathcal{M}_s(\Omega_{n-1} \times S^\prime)
       	\\
        & = 
        \Proba^\mathcal{M}_s(\neg\alpha \ltlUntil s) \times
		\Proba^\mathcal{M}_s(\Omega_{n-1} \times S^\prime)
	\end{align*}
    
    \begin{align*}
    	\text{(b)~} \Proba^\mathcal{M}_s(\ltlExists \alpha) &
    	= \Sigma_{n = 1}^{+\infty} \Proba^\mathcal{M}_s(\Omega_n \times S^\prime)
        \\
        & = 
        \frac{
        	\Proba^\mathcal{M}_s(\Omega_1 \times S^\prime)}{
        	1 - \Proba^\mathcal{M}_s(\neg\alpha \ltlUntil s)}
        \\
        & = 
        \frac{
        	\Proba^\mathcal{M}_s(\neg s \ltlUntil \alpha)}{
        	1 - \Proba^\mathcal{M}_s(\neg\alpha \ltlUntil s)}
	\end{align*}
    \qed
\end{proof}

\begin{lemma}\label{lem:min_max_reachability}
	Let $\mathcal{I} = (S,s_0,P,V)$ be an {\imc},
	$\mathcal{M}$ be an {\mc} satisfying $\mathcal{I}$, and
	$\alpha \subseteq A$ be a proposition.
	There exist {\mc}s $\mathcal{M}_1$ and $\mathcal{M}_2$ satisfying $\mathcal{I}$ with the same structure than $\mathcal{I}$
	 such that
	$\Proba^{\mathcal{M}_1}(\ltlExists \alpha) \leq \Proba^{\mathcal{M}}(\ltlExists \alpha) \leq
	\Proba^{\mathcal{M}_2}(\ltlExists \alpha)$.
\end{lemma}

Lemma~\ref{lem:min_max_reachability} is a consequence of Corollary \ref{lem:1sat_same_proba} and Lemma \ref{lem:reach_with_less_states} and
states that the maximal and the minimal probability of reaching a given proposition
is realized by Markov chains with the same structure than the {\imc}.

\begin{proof}[Lemma~\ref{lem:min_max_reachability}]
Let $\mathcal{I}$ be an {\imc} and $\mathcal{M}$ be a {\mc} satisfying $\mathcal{I}$.
Consider the sequence of {\mc}s $(\mathcal{M}_n)_{n \in \Nset}$ s.t.
$\mathcal{M}_0$ is the {\mc} satisfying $\mathcal{I}$ with degree $1$ obtained by Corollary~\ref{lem:1sat_same_proba}
and for all $n \in \Nset$, $\mathcal{M}_{n+1}$ is the {\mc} satisfying $\mathcal{I}$ with strictly less states than $\mathcal{M}_{n}$ and verifying 
$\Proba^{\mathcal{M}_{n+1}}(\ltlExists \alpha) \leq \Proba^{\mathcal{M}_{n}}(\ltlExists \alpha)$ 
given by Lemma~\ref{lem:reach_with_less_states} if
$\mathcal{M}_{n}$ does not have the same structure than $\mathcal{I}$ 
and equal to $\mathcal{M}_{n}$ otherwise.
By construction $(\mathcal{M}_n)_{n \in \Nset}$ is finite and its last element
is a Markov chain $\mathcal{M}^\prime$ with the same structure than $\mathcal{I}$
s.t. $\Proba^{\mathcal{M}^\prime}(\ltlExists \alpha) \leq \Proba^{\mathcal{M}}(\ltlExists \alpha)$.\qed
\end{proof}

\begin{paragraph}{\textnormal{\textbf{Theorem~\ref{thm:reachability-semantics-equivalence-imcs}}}}
	Let $\mathcal{I} = (S,s_0,P,V)$ be an {\imc}, $\alpha
        \subseteq A$ be a state label, ${\sim} \in \{\le,<,>,\ge\}$
        and $0<p<1$. $\mathcal{I}$ satisfies
        $\mathbb{P}^{\mathcal{I}}(\ltlExists \alpha) {\sim} p$ with
        the once-and-for-all semantics iff $\mathcal{I}$ satisfies
        $\mathbb{P}^{\mathcal{I}}(\ltlExists \alpha) {\sim} p$ with
        the at-every-step semantics.

\end{paragraph}

\begin{proof}
	Let $\mathcal{I} = (S,s_0,P,V)$ be an {\imc}, 
    $\alpha \subseteq A$ be a state label, 
    ${\sim} \in \{\le,<,>,\ge\}$ and $0 < p < 1$.
    Recall that according to an {\imc} satisfaction relation
    the property $\mathbb{P}^{\mathcal{I}}(\ltlExists \alpha) {\sim} p$ holds
    iff there exists an {\mc} $\mathcal{M}$ satisfying $\mathcal{I}$ 
    (with the chosen semantics) such that
	$\mathbb{P}^{\mathcal{M}}(\ltlExists \alpha) {\sim} p$.
    Recall also that $\satisfactionImc$ is more general than $\satisfactionImcOnce$: 
    for all {\mc} $\mathcal{M}$ if $\mathcal{M} \satisfactionImcOnce \mathcal{I}$ then 
    $\mathcal{M} \satisfactionImcOnce \mathcal{I}$
    \begin{itemize}
    	\item[$\lbrack\Rightarrow\rbrack$] Direct from the fact that 
        $\satisfactionImcOnce$ is more general than $\satisfactionImcOnce$ 
        (\cf Appendix~\ref{ap:compare_imcs_satisfaction_relations})
        
        \item[$\lbrack\Leftarrow\rbrack$] 
        $\mathbb{P}^{\mathcal{I}}(\ltlExists \alpha) {\sim} p$ with
        the at-every-step semantics implies that there exists
        a {\mc} $\mathcal{M}$ s.t.
        $\mathcal{M} \satisfactionImc \mathcal{I}$ and $\mathcal{M} {\sim} p$.
        Thus by Lemma~\ref{lem:min_max_reachability} there exists a {\mc} $\mathcal{M}^\prime$ s.t.
        $\mathcal{M}^\prime \satisfactionImc \mathcal{I}$, $\mathcal{M}^\prime {\sim} p$,
        $\mathcal{M}^\prime$ has the same structure than $\mathcal{I}$.
        However if $\mathcal{M}^\prime$ has the same structure than $\mathcal{I}$
        then $\mathcal{M}^\prime$ satisfies $\mathcal{I}$ with the one-and-for-all semantics.
	\end{itemize}\qed
\end{proof}

\subsection{{\csp} Encodings for Quantitative Reachability}
 \begin{proposition}\label{prop:model_reach_label}
 	Let  $\mathcal{P} = (S, s_0 , P, V, Y)$ be a {\pimc} and
 	$\alpha \subseteq A$ be a state label.
 	There exists a {\mc} $\mathcal{M} \satisfactionPimc \mathcal{P}$ iff
 	there exists a valuation $v$ solution of the \csp\ $\MerPrime(\mathcal{P},\alpha)$
 	s.t. for each state $s \in S$: 
 	$v(\lambda_s)$ is equals to {\true} iff $\Proba^\mathcal{M}_s(\ltlExists \alpha) \neq 0$.
 \end{proposition}

Let  $\mathcal{P} = (S, s_0 , P, V, Y)$ be a {\pimc} and
$\alpha \subseteq A$ be a state label.
For each state $s \in S$, the variable $\alpha_s$ plays a symmetric role to 
the variable $\omega_s$ from $\Mer$: instead of indicating the existence
of a path from $s_0$ to $s$, it characterizes the existence of a path
from $s$ to a state labeled with $\alpha$. 
As for the $\Mer$ {\csp} encoding, each solution of the {\csp} $\Mer(P, \alpha)$
corresponds to a {\mc} $\mathcal{M}$ satisfying $\mathcal{P}$ w.r.t. $\satisfactionPimcOnce$.
Furthermore the constraints added in $\MerPrime$ ensures that $\alpha_s$ is equal to $k$
iff there exists a path of length $k-1$ with non zero probability from $s$ to a state label with $\alpha$ in $\mathcal{M}$.
Thus by Constraint~\ref{encoding_erprime_bool_var}, 
variables $\lambda_s$ is equal to {\true} iff
there exists a path with non zero probability from the initial state
$s_0$ to a state labeled with $\alpha$ passing by $s$.

 \begin{proposition}\label{prop:model_quant_reachability}
 	Let  $\mathcal{P} = (S, s_0 , P, V, Y)$ be a {\pimc} and
 	$\alpha \subseteq A$ be a proposition.
 	There exists a {\mc} $\mathcal{M} \satisfactionPimc \mathcal{P}$
 	iff
 	there exists a valuation $v$ solution of the {\csp} $\MerExt(\mathcal{P},\alpha)$ s.t.
 	$v(\pi_{s})$ is equal to $\Proba^\mathcal{M}_s(\ltlExists \alpha)$ if $s$ is reachable from the initial state $s_0$ in $\mathcal{M}$
 	and is equal to $0$ otherwise.
 \end{proposition}

Let  $\mathcal{P} = (S, s_0 , P, V, Y)$ be a {\pimc} and
$\alpha \subseteq A$ be a state label.
$\MerExt$ extends the {\csp} $\MerPrime$ that produces a 
$\mc$ $\mathcal{M}$ satisfying $\mathcal{P}$ (\cf Proposition~\ref{prop:model_reach_label})
by computing the probability of reaching $\alpha$ in $\mathcal{M}$.
In order to compute this probability, we
use standard techniques from~\cite{Baier2008PMC} that require the
partitioning of the state space into three sets $S_{\top}$,
$S_{\bot}$, and $S_?$ that correspond to states reaching
$\alpha$ with probability $1$, states from which $\alpha$ cannot be
reached, and the remaining states, respectively. Once this partition is chosen, the
reachability probabilities of all states in $S_?$ are computed as the
unique solution of an equation system (see~\cite{Baier2008PMC},
Theorem 10.19, p.766). 
Recall that for each state $s \in S$
variable $\alpha_s$ is equal to {\true} iff $s$ is reachable in $\mathcal{M}$ and $s$ can reach $\alpha$ with a non zero probability.
Thus we consider $S_\bot = \{s \ |\ \alpha_s = \false\}$,
$S_\top = \{s \ |\ V(s) = \alpha\}$, and
$S_? = S \setminus (S_\bot \cup S_\top)$.
Finally constraints in $\MerExt$ encodes the equation system from \cite{Baier2008PMC}
according to chosen $S_\bot$, $S_\top$, and $S_?$.
Thus $\pi_{s_0}$ is equal to the probability in $\mathcal{M}$
to reach $\alpha$.

\begin{paragraph}{\textnormal{\textbf{Theorem~\ref{thm:pimc_reachability_in_cp}.}}}
	Let $\mathcal{P} = (S, s_0 , P, V, Y)$ be a \pimc,
	$\alpha \subseteq A$ be a label,
	$p \in [0, 1]$,
    ${\sim} \in \{\leq,<, \geq,>\}$ be a comparison operator,
    and $(X,D,C)$ be $\MerExt(\mathcal{P}, \alpha)$:
    \vspace*{-0.05cm}
    \begin{itemize}
    	\item 
			\csp\ $(X,D,C \cup (\pi_{s_0} \sim p))$
    		is satisfiable iff 
    		$\exists \mathcal{M} \satisfactionPimc \mathcal{P}$ s.t. $\Proba^\mathcal{M}(\ltlExists \alpha) \sim p$
   		\item
			\csp\ $(X,D,C \cup (\pi_{s_0} \not\sim p))$
    		is unsatisfiable iff 
    		$\forall \mathcal{M} \satisfactionPimc \mathcal{P}$: $\Proba^\mathcal{M}(\ltlExists \alpha) \sim p$
	\end{itemize}
\end{paragraph}

Let  $\mathcal{P} = (S, s_0 , P, V, Y)$ be a {\pimc},
$\alpha \subseteq A$ be a state label,
$p \in [0, 1]$, and
${\sim} \in \{\leq,<, \geq,>\}$ be a comparison operator.
Recall that $\MerExt(\mathcal{P},\alpha)$ is a {\csp} s.t.
each solution corresponds to a $\mc$ $\mathcal{M}$ satisfying $\mathcal{P}$ where $\pi_{s_0}$ is equal to $\Proba^\mathcal{M}(\ltlExists \alpha)$.
Thus adding the constraint $\pi_{s_0} \sim p$ allows to find a {\mc}
$\mathcal{M}$ satisfying $\mathcal{P}$ s.t. $\Proba^\mathcal{M}(\ltlExists \alpha) \sim p$.
This concludes the first item presented in the theorem.
For the second item, we use Theorem~\ref{thm:reachability-semantics-equivalence-imcs} with the Proposition~\ref{prop:model_quant_reachability} which ensure that if the {\csp} $\MerExt(\mathcal{P},\alpha)$ plus the constraint $\pi_{s_0} \not\sim p$ is not satisfiable then 
there is no {\mc} satisfying $\mathcal{P}$ w.r.t. $\satisfactionPimc$ s.t. $\Proba^\mathcal{M}(\ltlExists \alpha) \not\sim p$;
thus $\Proba^\mathcal{M}(\ltlExists \alpha) \sim p$ 
for all {\mc} satisfying $\mathcal{P}$ w.r.t. $\satisfactionPimc$.

\end{document}